\documentclass{article}
\usepackage{epsfig,amssymb,float,psfig}
\input{epsf}
\newcommand{\be}{\begin{equation}}
\newcommand{\ee}{\end{equation}}
\newcommand{\ba}{\begin{eqnarray}}
\newcommand{\ea}{\end{eqnarray}}
\newcommand{\beq}{\begin{equation}}
\newcommand{\eeq}{\end{equation}}
\newcommand{\beqa}{\begin{eqnarray}}
\newcommand{\eeqa}{\end{eqnarray}}

\newcommand{\vs}{\vspace{-0.15cm}}

\begin{document}

\begin{flushright}
{\tiny  HISKP-TH-05/01}\\
{\tiny  FZJ-IKP(TH)-2005-01} \\
{\tiny  $\,$} \\
\end{flushright}

\vspace{1cm}

\begin{center}

\bigskip

{{\Large\bf Baryon masses, chiral extrapolations, and all that\footnote{This
research is part of the EU Integrated Infrastructure Initiative Hadron 
Physics Project under contract number RII3-CT-2004-506078. 
Work supported in part by DFG (SFB/TR 16 ``Subnuclear Structure of Matter'').}
}}

\end{center}

\vspace{.3in}

\begin{center}
{\large
Matthias Frink$^{\dagger,}$\footnote{email: mfrink@itkp.uni-bonn.de},
Ulf-G. Mei{\ss}ner$^{\dagger,\ddagger,}$\footnote{email:
  meissner@itkp.uni-bonn.de}, 
Ilka Scheller$^{\dagger,}$\footnote{email: scheller@itkp.uni-bonn.de}
}

\vspace{1cm}

$^\dagger${\it Helmholtz-Institut f\"ur Strahlen- und Kernphysik (Theorie),
                  Universit\"at  Bonn\\
                  Nu{\ss}allee 14-16, D-53115 Bonn, Germany}

\bigskip

$^\ddagger${\it Forschungszentrum J\"ulich, Institut f\"ur Kernphysik
(Theorie)\\ D-52425 J\"ulich, Germany}

\bigskip

\end{center}

\vspace{.6in}

\thispagestyle{empty}

\begin{abstract}\noindent
We calculate the baryon octet masses to fourth order in chiral perturbation theory
employing dimensional and cut-off regularization. We analyze the pion and kaon
mass dependences of the baryon masses based on the MILC data. We show that 
chiral perturbation theory gives stable chiral extrapolation functions for 
pion (kaon) masses below 550~(600)~MeV. The pion-nucleon sigma term in SU(3) is
also investigated, we find $\sigma_{\pi N} (0) = 50.7 \ldots 53.7\,$MeV.
\end{abstract}

\vfill

\pagebreak

\section{Introduction and summary}
\def\theequation{\arabic{section}.\arabic{equation}}
\setcounter{equation}{0}
\label{sec:intro}

The masses of the ground state baryon octet are of fundamental
importance in the investigation of three-quark states in QCD.
With the advent of improved techniques in lattice QCD and systematic
studies within the framework of chiral perturbation theory, one can hope
to gain an understanding of these quantities from first principles. 
Present day lattice calculations are done at unphysical quark masses
above the physical values, therefore chiral extrapolations are
needed to connect lattice results with the physical world, provided that
the masses are not too high (for an early approach to this problem,
see e.g. \cite{Aus}).  With the recent  data
from the MILC collaboration \cite{MILC,MILCnew} it appears to be possible
to apply chiral extrapolation functions derived from chiral perturbation
theory (CHPT) (for a review, see \cite{BKMrev}). 
In this paper, we analyze the baryon masses as a function of the
pion and the kaon masses in  CHPT based on the MILC data. 
Other pertinent lattice papers are e.g. \cite{TXL,Getal,MTc,UKQCD,CPPACS}.
It should be
said from the beginning that we ignore the effects of the a) the finite
volume, b) the finite lattice spacing and c) the staggered approximation
in this study\footnote{Some of these effects are studied in 
\cite{QCDSFUKQCD,paulo,tib} (and references therein).} 
since our aim is more modest - we want to find out whether
these chiral extrapolations can be used for the presently available
lattice data. Once this test is performed, one should then apply
the full formalism including the abovementioned effects. 
Note that the quark mass expansion of QCD is turned
into an expansion in Goldstone boson (GB) masses in CHPT - we thus use
both terms synonymously.

\begin{figure}[H]
\vspace{-0.2cm}
\parbox{.50\textwidth}{\epsfig{file= 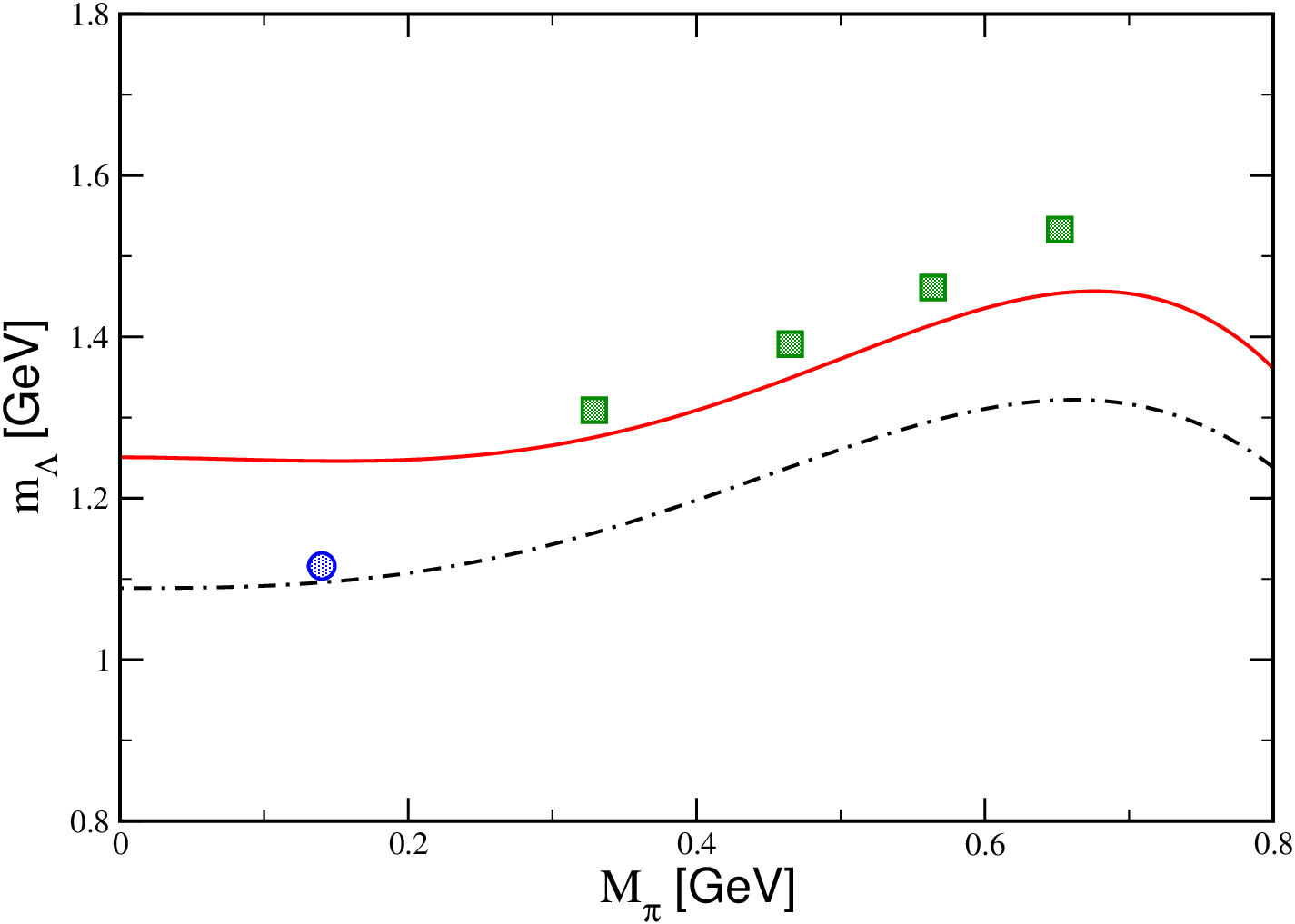,width=.44\textwidth,angle=0,
silent=,clip=}}
\hfill
\parbox{.50\textwidth}{\epsfig{file= 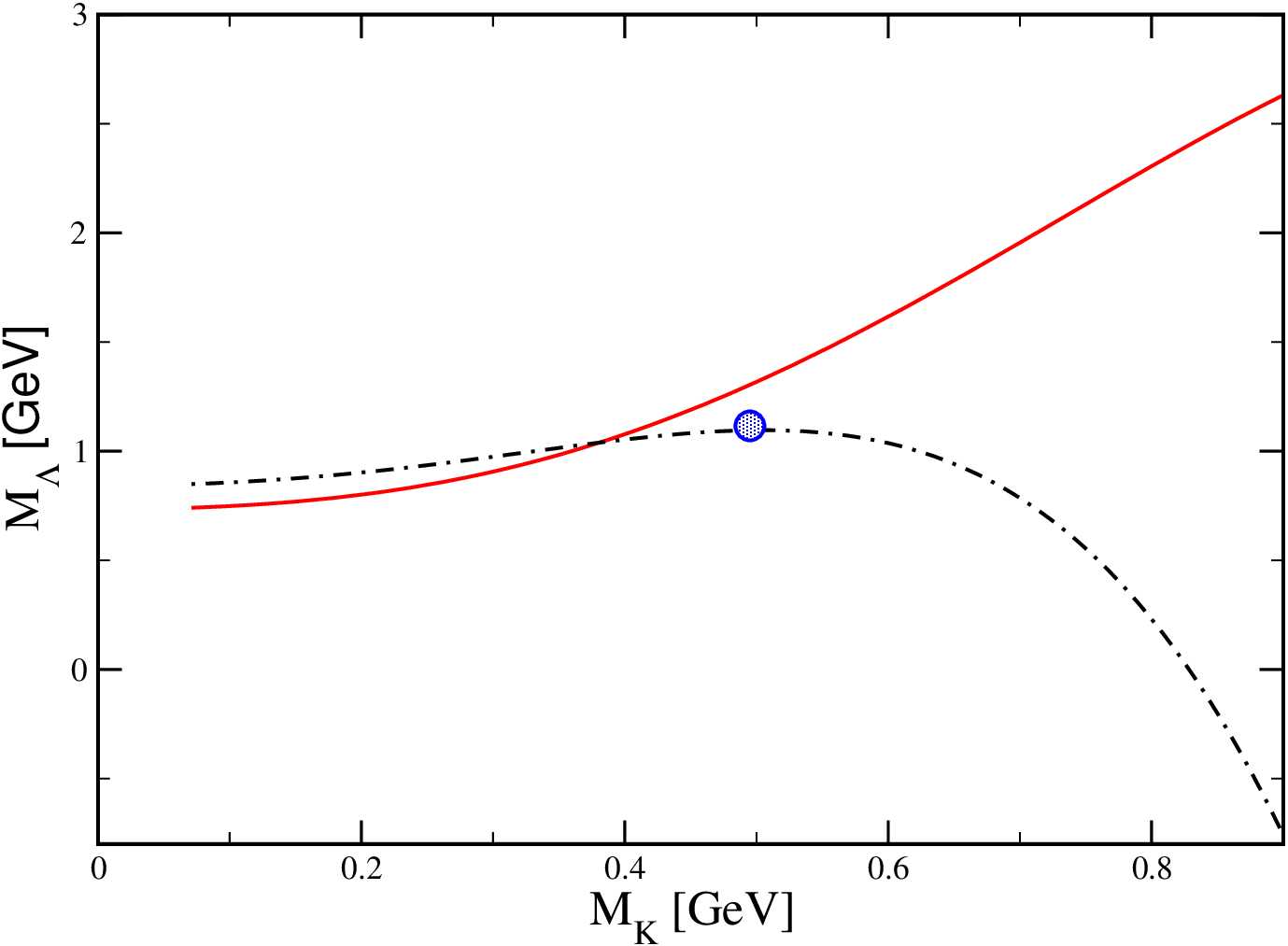,width=.44\textwidth,angle=0,
silent=,clip=}}
\vspace{-0.3cm}
\begin{center}
\caption{Left panel: Pion mass dependence of the $\Lambda$ mass for various
sets of the LECs $d_i$ as explained in the text. Right panel: Kaon mass
dependence.
\label{fig:hypL}}
\end{center}
\end{figure}

\begin{figure}[H]
\vspace{-0.4cm}
\parbox{.50\textwidth}{\epsfig{file= 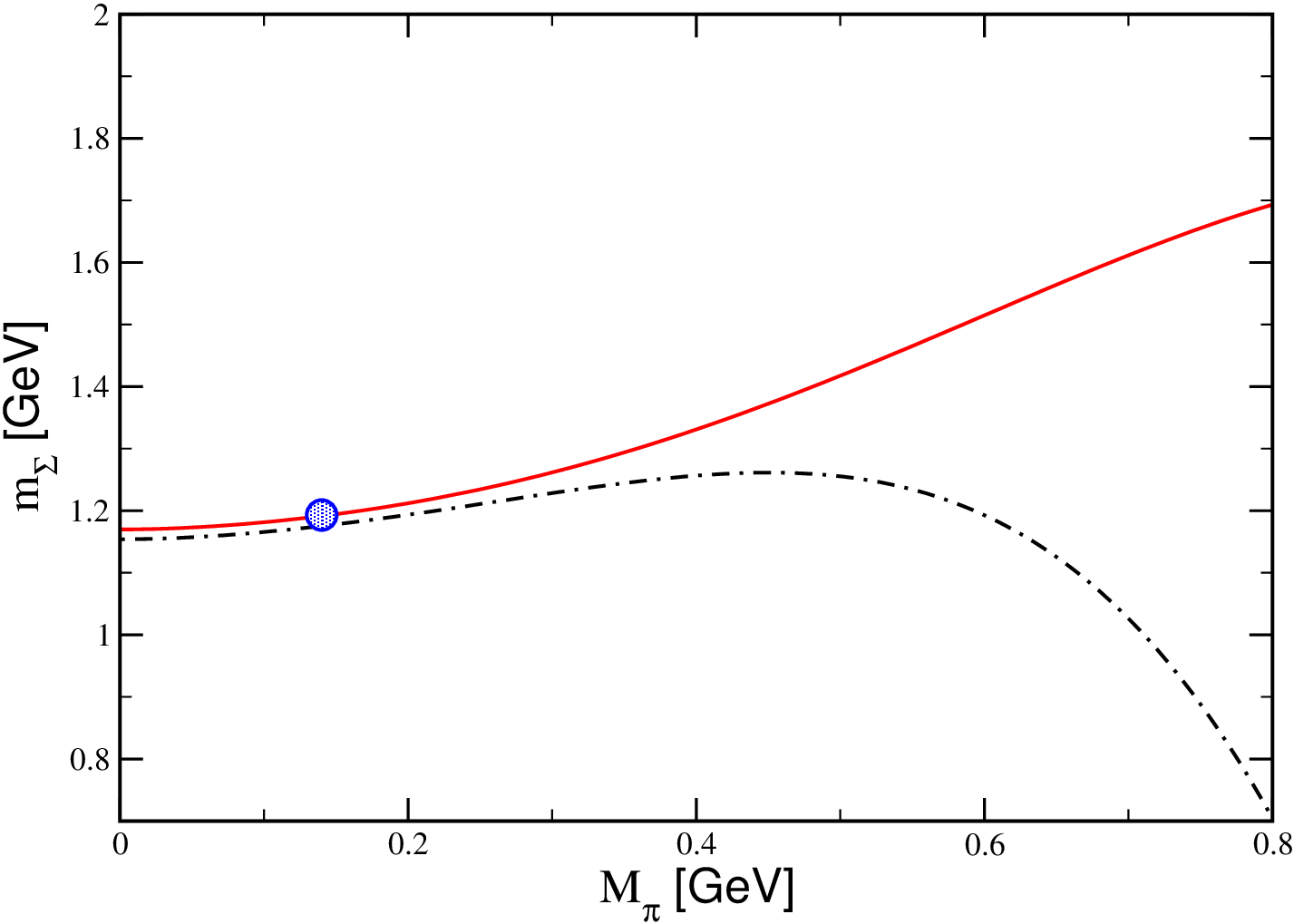,width=.44\textwidth,angle=0,
silent=,clip=}}
\hfill
\parbox{.50\textwidth}{\epsfig{file= 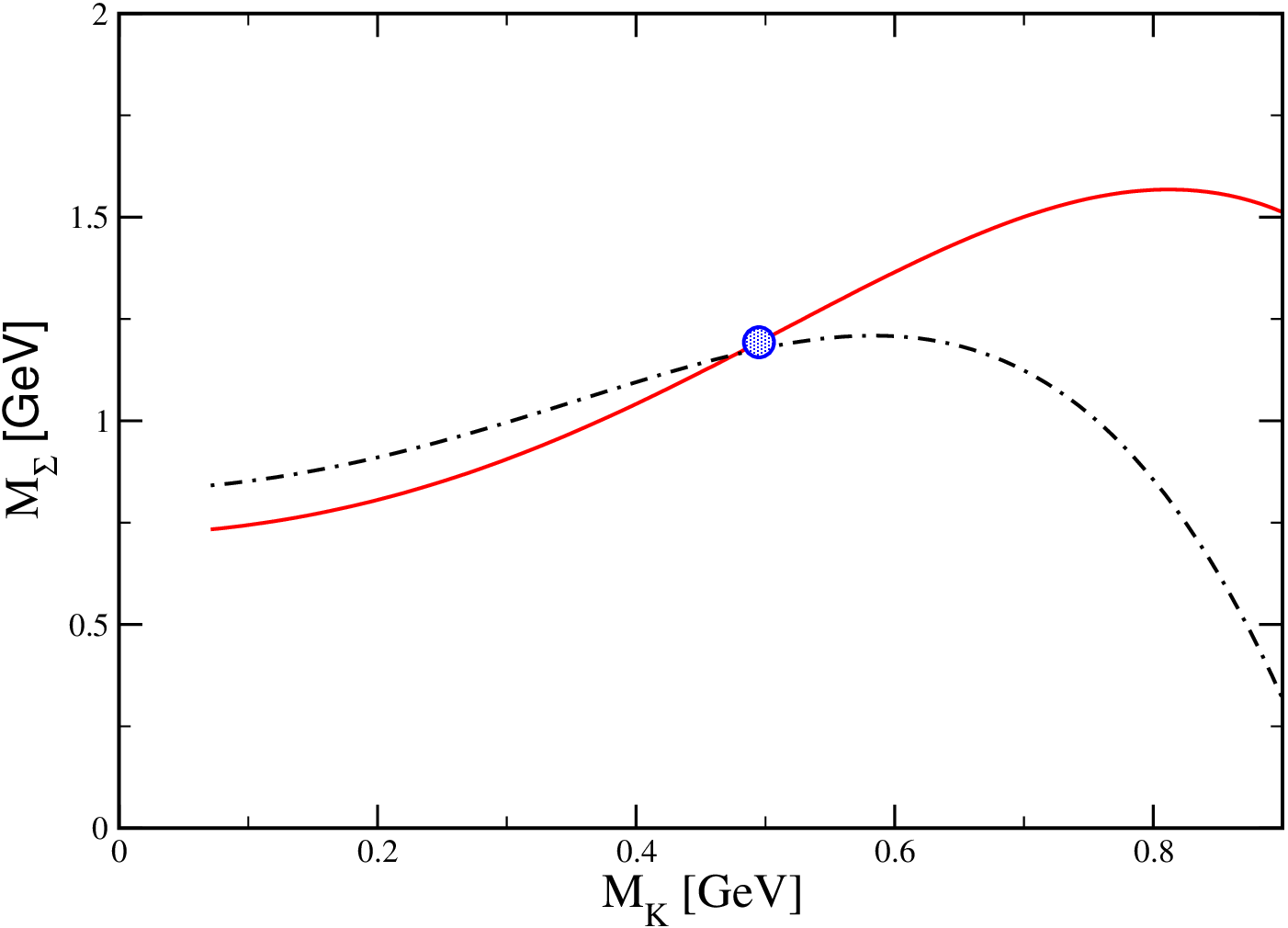,width=.44\textwidth,angle=0,
silent=,clip=}}
\vspace{-0.3cm}
\begin{center}
\caption{Left panel: Pion mass dependence of the $\Sigma$ mass for various
sets of the LECs $d_i$ as explained in the text. Right panel: Kaon mass
dependence.
\label{fig:hypS}}
\end{center}
\end{figure}

\begin{figure}[H]
\vspace{-0.4cm}
\parbox{.50\textwidth}{\epsfig{file= 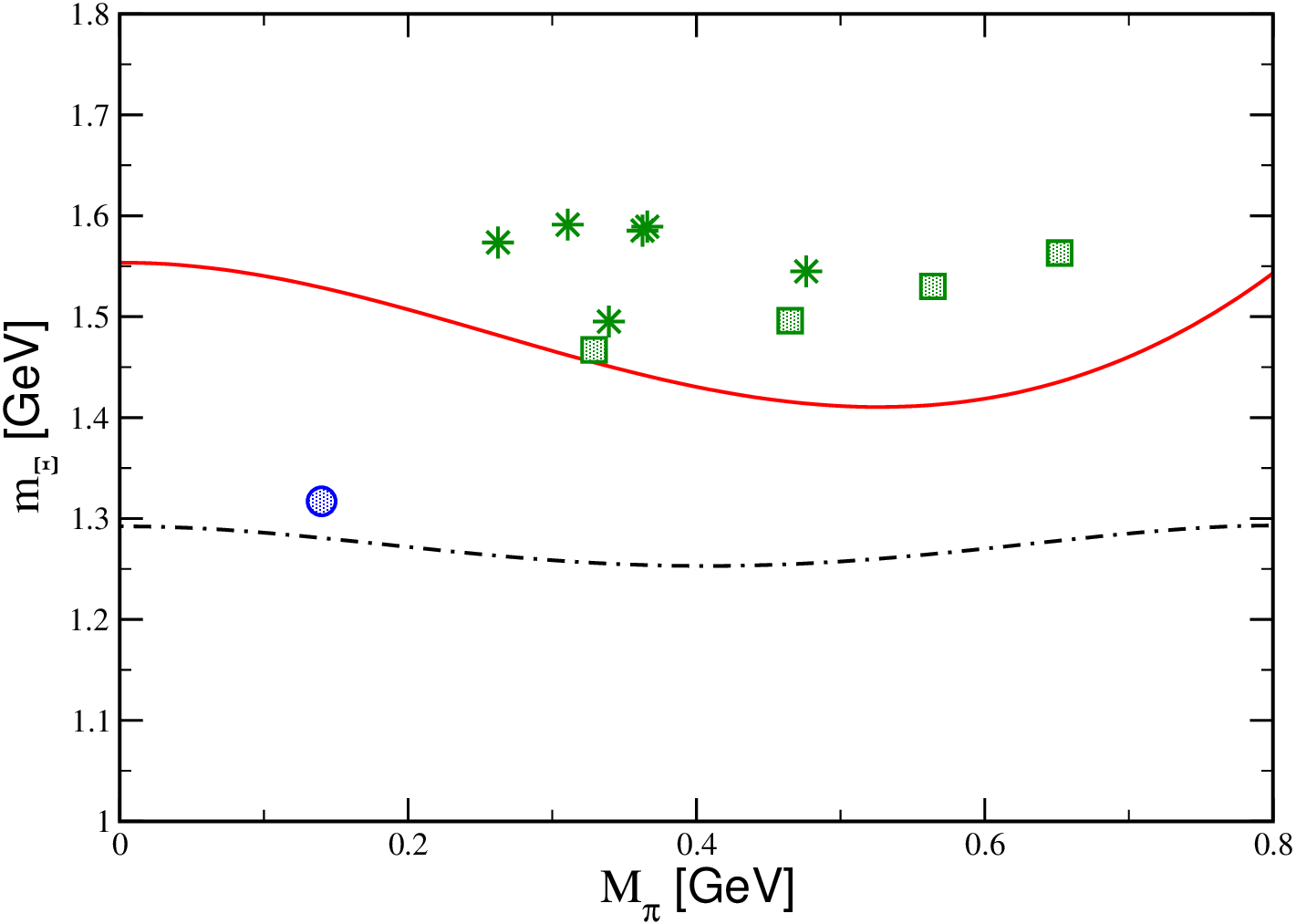,width=.44\textwidth,angle=0,
silent=,clip=}}
\hfill
\parbox{.50\textwidth}{\epsfig{file= 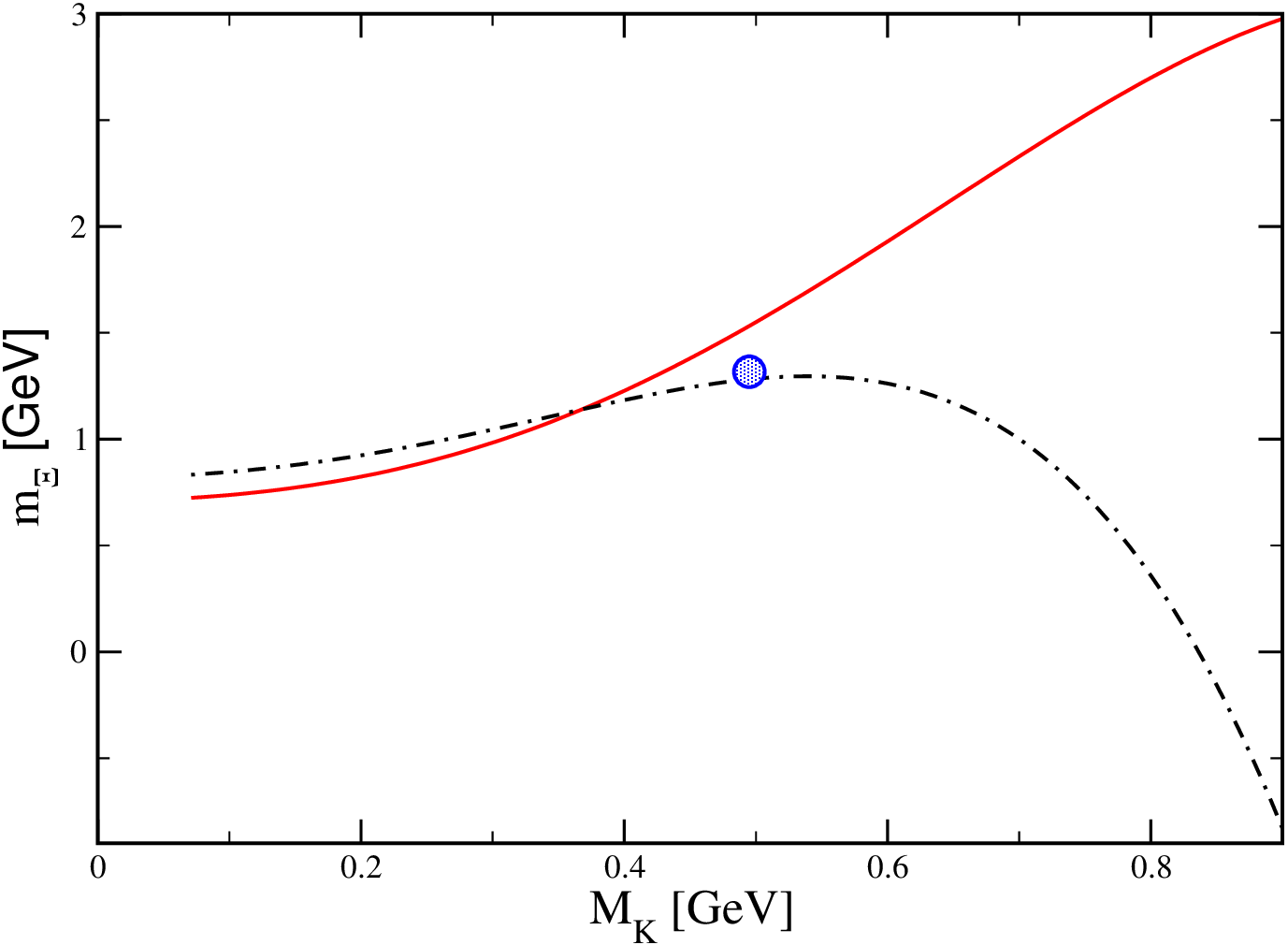,width=.44\textwidth,angle=0,
silent=,clip=}}
\vspace{-0.3cm}
\begin{center}
\caption{Left panel: Pion mass dependence of the $\Xi$ mass for various
sets of the LECs $d_i$ as explained in the text. Right panel: Kaon mass
dependence.
\label{fig:hypX}}
\end{center}
\end{figure}

\medskip\noindent
The baryon masses have been analyzed in various versions of baryon
CHPT to third and fourth order, for an incomplete list of references
see \cite{EJ,BKM,BM,Tori,FM,AWL,Mzmasses}. Our investigation extends the
work in the two-flavor sector presented in \cite{BHMcut} and we
heavily borrow from the earlier SU(3) calculations of \cite{BM,FM}.
The pertinent results of this investigation can be summarized as
follows:

\begin{itemize}
\item[1)]We have calculated the baryon masses to third and fourth order
in the chiral expansion making use of cut-off regularization as proposed
in \cite{BHMcut}. As in that paper, we have also considered an improvement
term at third order to cancel the leading cut-off dependence in the
baryon masses, see Sect.~\ref{sec:mass3imp}.
\item[2)]The improvement term consists of three independent terms, whose
cut-off independent coefficients have been determined by considering the
nucleon mass (to allow for a direct comparison with the SU(2) calculation
of Ref.~\cite{BHMcut}). We have demanded that for the physical pion and the 
physical kaon mass the nucleon mass passes through its physical value.
This fixes two parameter combinations.
The third parameter is determined from a best fit to the trend of
the earlier MILC data \cite{MILC} for $m_N (M_\pi)$, cf. Fig.~\ref{fig:mN3imp}, under the
condition that the deviation from the earlier determination of the
corresponding low-energy constants in \cite{BM} 
is of natural size. We find indeed
a visible improvement in the description of the lattice data and also 
much better stability under variations of the cut-off.
\item[3)]The full fourth-order calculation utilizing the low-energy constants
as determined from the improvement term leads to an accurate description of
the MILC data for pion masses below 550~MeV, see again
Fig.~\ref{fig:mN3imp}.  Note that the two lowest mass points of the more recent
MILC data \cite{MILCnew} can not be well described. Also,
the use of the low-energy constants (LECs)
from \cite{BM} lead to a less satisfactory description. We have also discussed the
theoretical uncertainty of this procedure, cf. Fig.~\ref{fig:mN4}.
\item[4)]From the pion mass dependence of the nucleon mass, we can deduce the
pion-nucleon sigma term. For the best sets of low-energy constants, we find
$\sigma_{\pi N} (0) = 50.7 \ldots 53.7\,$MeV.
\item[5)]The kaon mass dependence of the nucleon mass is less well determined.
Still, the extrapolation functions can be applied to kaon masses below
$\simeq 600\,$MeV.
We deduce that the baryon  octet mass in the chiral limit lies in the
interval $710\,{\rm MeV} \lesssim m_0 \lesssim 1070\,{\rm MeV}$. This is
consistent with earlier estimates, see e.g. \cite{BM}.
\item[6)]We have also considered the pion and kaon mass dependences of the
$\Lambda$, the $\Sigma$ and the $\Xi$ and compared to the existing MILC
data, cf. Figs.~\ref{fig:hypL},\ref{fig:hypS},\ref{fig:hypX}. 
Note that we have not fitted to these masses.  Our chiral extrapolations 
for the $\Sigma$ and in particular for the $\Xi$ as a function of the 
pion mass are flatter than the MILC data. This is partly due to our strategy
of fixing all parameters on the nucleon mass. We remark, however, that one
should expect a decreased pion mass dependence as the number of strange
valence quarks increases.
\end{itemize}

\medskip\noindent
The material in this paper is organized as follows. Section~\ref{sec:formalism}
contains the effective Lagrangian and a short discussion about the various
regularization methods employed in our calculations. In
section~\ref{sec:masses}, the ground state  baryon octet masses are
given at third, third improved and fourth order in the chiral expansion.
In particular, we concentrate on the differences to the SU(2) case
\cite{BHMcut} (we refer to that paper for many details).
Our results for the various baryon masses as functions of the pion and
the kaon masses and the stability of these results under cut-off variations
are given and discussed in  section~\ref{sec:res}. Many technicalities are 
relegated to the appendices.


\section{Formalism I: Generalities}
\def\theequation{\arabic{section}.\arabic{equation}}
\setcounter{equation}{0}
\label{sec:formalism}

In this section, we display the effective Lagrangian underlying our
calculations and discuss briefly the cut-off regularization utilized and
its relation to the more standard dimensional regularization (DR). We
borrow heavily from the work presented in Refs.\cite{BM,BHMcut} and refer
the reader for more details to these papers.

\subsection{Effective Lagrangian}
\label{sec:Leff}

Our calculations are based on an  effective chiral meson-baryon Lagrangian
in the presence of external sources (like e.g. photons)  supplemented
by a power counting in terms of quark (meson) masses and small external momenta.
Its generic form consists of a string of terms with increasing chiral dimension,
\begin{eqnarray}\label{L}
{\cal L} = {\cal L}_{\phi B}^{(1)}+{\cal L}_{\phi B}^{(2)} 
+ {\cal L}_{\phi B}^{(3)} +{\cal L}_{\phi B}^{(4)} + {\cal L}_\phi^{(2)} 
+ {\cal L}_\phi^{(4)}~.
\end{eqnarray}
Here, $B$ collects the baryon octet and $\phi$ stands for the Goldstone
boson octet. The superscript denotes the power in the genuine small parameter $q$ 
(denoting Goldstone boson masses and/or external momenta). 
The explicit representations of  $\phi$ and $B$ are:
\begin{eqnarray}
 \phi(x)=\left(\begin{array}{ccc} \pi^0+\frac{1}{\sqrt{3}}\eta &  
\sqrt{2}\pi^+ & \sqrt{2}K^+ \\ 
\sqrt{2}\pi^- & -\pi^0+\frac{1}{\sqrt{3}}\eta & \sqrt{2} K^0 \\ 
\sqrt{2} K^- & \sqrt{2} \bar{K}^0 & -\frac{2}{\sqrt{3}}\eta 
\end{array}\right)~, \quad
B (x) =\left(\begin{array}{ccc} 
\frac{1}{\sqrt{2}}\Sigma^0+\frac{1}{\sqrt{6}}\Lambda & \Sigma^+ & p\\
\Sigma^- & -\frac{1}{\sqrt{2}}\Sigma^0+\frac{1}{\sqrt{6}}\Lambda & n\\
\Xi^- & \Xi^0 & -\frac{2}{\sqrt{6}}\Lambda \end{array}\right)~.
\end{eqnarray}
A complete one--loop (fourth
order) calculation must include all tree level graphs with insertions
from all terms given in Eq.~(\ref{L}) and loop graphs with at most
one insertion from ${\cal L}_{\phi B}^{(2)}$ or  ${\cal L}_{\phi}^{(2)}$.  
Throughout, we employ the 
heavy baryon approach, which allows for a consistent power counting since the
large mass scale (the baryon mass $m_B$) is transformed from the propagator into 
a string of $1/m_B$ suppressed interaction vertices. The lowest order
(dimension one) effective Lagrangian takes the canonical form
\begin{eqnarray}
 {\cal L}_{\phi B}^{(1)} & =& i\, \mbox{Tr}(\bar{B}[v\cdot D, B]) 
+ F\, \mbox{Tr}(\bar{B} S_\mu [u^\mu,B]) 
+ D \, \mbox{Tr}(\bar{B} S_\mu\{u^\mu,B\})~,
\end{eqnarray}
where $v_\mu$ is the four-velocity of the baryon subject to the constraint
$v^2 =1$, Tr denotes the trace in flavor space and $D$ and $F$ are the leading
axial-vector couplings, $D\simeq 3/4$ and $F \simeq 1/2$. Furthermore, $S_\mu$
is the  spin-vector and $u_\mu = i u^\dagger \nabla_\mu U u^\dagger$, where
$U = u^2$ collects the Goldstone bosons (for more details, see \cite{BKMrev}).  
For the calculation of the self-energy (mass), it suffices to use the 
partial derivative
$\partial_\mu$ instead of the chiral covariant derivative $D_\mu$. The
dimension two chiral Lagrangian can be decomposed as (for details see
\cite{BM} and \cite{FM})
\begin{eqnarray}
{\cal L}_{\phi B}^{(2)}&=& {\cal L}_{\phi B}^{(2,\rm{br})} 
+ \sum_{i=1}^{19} b_i O_i^{(2)} + {\cal L}_{\phi B}^{(2,\rm{rc})}
\end{eqnarray}
with
\begin{eqnarray}
 {\cal L}_{\phi B}^{(2,\rm{br})} & = & b_D \mbox{Tr}[\bar{B}\{\chi_+,B\}] 
  + b_F \mbox{Tr}[\bar{B}[\chi_+,B]] + b_0 \mbox{Tr}[\bar{B}B]\mbox{Tr}[\chi_+],\\
  \sum_{i=1}^{19} b_i O_i^{(2)} & = & b_1 \mbox{Tr}[\bar{B}[u_\mu,[u^\mu,B]]] 
  + b_2 \mbox{Tr}[\bar{B}[u_\mu,\{u^\mu,B\}]]
   + b_3 \mbox{Tr}[\bar{B}\{u_\mu,\{ u^\mu,B\}\}]\nonumber\\
            &+& \bigg((b_4- m_0 b_{15})m_0 +\frac{1}{4}(b_{12}-m_0 
            b_{18})m_0\bigg)\mbox{Tr}[\bar{B}[v\cdot u,[v \cdot u,B]]]\nonumber\\
           &+& (b_5+b_6-m_0 b_{16}) m_0\mbox{Tr}[\bar{B}[v\cdot u,\{v\cdot u, B\}]]\nonumber\\
           &+& \bigg((b_7-m_0 b_{17})m_0+\frac{3}{4}(b_{12}-m_0 b_{18})\bigg)m_0
           \mbox{Tr}[\bar{B}\{ v\cdot u,\{ v\cdot u,B\}\}]\nonumber\\
          &+& b_8\mbox{Tr}[\bar{B}B]\mbox{Tr}[u^\mu u_\mu]
          +b_{11} 2 i\epsilon^{\mu\nu\alpha\beta}\mbox{Tr}[\bar{B}u_\mu]v_\alpha 
               S_\beta\mbox{Tr}[u_\nu B]\nonumber\\
          &+& \bigg((b_9-m_0b_{19})m_0-\frac{1}{2}(b_{12}-m_0 b_{18})m_0\bigg)
          \mbox{Tr}[\bar{B}B]\mbox{Tr}[v\cdot u\; v\cdot u]\nonumber\\
&+& b_{13} 2 i \epsilon^{\mu\nu\alpha\beta}\mbox{Tr}[\bar{B}v_\alpha S_\beta
      [[u_\mu,u_\nu],B]]
 + b_{14} 2 i \epsilon^{\mu\nu\alpha\beta}\mbox{Tr}[\bar{B}v_\alpha S_\beta 
        \{[u_\mu,u_\nu],B\}]~,
\end{eqnarray}
with $m_0$ the octet baryon mass in the chiral limit.
Explicit symmetry breaking embodied in the external source $\chi_+ \sim
 {\cal M}$ (with $\cal M$ the diagonal quark mass matrix)  only
starts at this order, collected in ${\cal L}_{\phi B}^{(2,\rm{br})}$. 
It is parameterized in terms of the LECs $b_0$, $b_D$ and
$b_F$. Throughout, we work in the isospin limit $m_u = m_d = \hat m$ and thus
consider four different baryon states, the nucleon doublet ($N$), 
the lambda ($\Lambda$),
the sigma triplet ($\Sigma$)   and the cascade doublet ($\Xi$) (Isospin
breaking corrections are discussed in \cite{FM} and \cite{TWL}).
The operators with the LECs $b_i$ ($i = 1, \ldots , 19)$ only appear
as insertions in fourth order tadpole graphs (see \cite{FM} for a detailed
discussion).
Note that in \cite{BM} the contributions from various
combinations of dimension two LECs 
were effectively subsumed
in one corresponding coupling since  operators $\sim k^2$ and 
$\sim (v\cdot k)^2$ lead to
the same contribution to the baryon masses. We can only do that at later stage
in the calculation so as to be able to consistently work out the
renormalization of these dimension two operators. As will become clear later,
while in DR the $b_i$ are finite numbers, this is not the case if one employs
cut-off regularization. Finally, we remark that the recoil terms collected in
$ {\cal L}_{\phi B}^{(2,\rm{rc})}$ are given in \cite{BM}. Similarly, there
are  further recoil corrections $\sim 1/m_B^2$ collected in 
$ {\cal L}_{\phi B}^{(3,\rm{rc})}$. These involve no unknown parameters, their
explicit form is also given in  \cite{BM}. To end this section, we give the
fourth order terms relevant for our calculations,
\begin{eqnarray} \label{L4}
{\cal L}_{\phi B}^{(4)} & = & d_1 \mbox{Tr}(\bar{B}[\chi_+,[\chi_+,B]])
+d_2\mbox{Tr}(\bar{B}[\chi_+,\{\chi_+,B\}])
                 +  d_3\mbox{Tr}(\bar{B}\{\chi_+,\{\chi_+,B\}\})
+d_4\mbox{Tr}(\bar{B}\chi_+)\mbox{Tr}(\chi_+B)\nonumber\\
                 &&  + \, d_5 \mbox{Tr}(\bar{B}[\chi_+,B])\mbox{Tr}(\chi_+)
+d_7 \mbox{Tr}(\bar{B}B)\mbox{Tr}(\chi_+)\mbox{Tr}(\chi_+)
                 +  d_8 \mbox{Tr}(\bar{B}B)\mbox{Tr}(\chi_+^2)~.
\end{eqnarray}
We remark that we have employed the notation of \cite{BM} to facilitate the
comparison with that work and also use some of the LECs determined there.

\subsection{Regularization schemes}
\label{sec:reg}

We briefly recall the salient features of the various regularization
schemes employed in calculating the baryon masses. Heavy baryon CHPT
together with DR was used e.g. in the early papers \cite{EJ,BKM,BM}
and a fourth order calculation using infrared regularization (which is
also based on DR to deal with the UV divergences in the loop graphs)
was reported in \cite{FM} (for an earlier incomplete calculation, see
\cite{Tori} and a recent calculation in the extended on-mass shell
renormalization scheme was reported in \cite{Mzmasses}). 
To be definite, consider the leading one-loop pion graph
for the nucleon mass  (the sunset diagram with insertions from the leading 
order Lagrangian which is of third order). In the heavy baryon approach, 
it is given by
\begin{equation}
I_{N}^\pi = \frac{c}{4}\, J(0) \, M_\pi^2~, \quad
J(0) = \frac{1}{i}\int\frac{d^d k}{(2\pi)^d}\frac{1}{(M_\pi^2-k^2-i\epsilon)
(v\cdot k-i\epsilon)}~,
\end{equation}
with $c= (D+F)^2/F_0^2$ and $F_0$ the pseudoscalar decay constant in the chiral
limit. In DR, the loop function $J(0)$ is finite and can be expressed as
 \begin{equation}\label{J}
J(0)= M_\pi^{d-3} (4\pi)^{-d/2} \,\Gamma\left(\frac{1}{2}\right)\,
\Gamma\left(\frac{3-d}{2}\right)
=-\frac{M_\pi}{8\pi}~,
\end{equation}
with $d$ the number of space-time dimensions and we have set $d = 4$ on
the right-hand-side of Eq.~(\ref{J}). This gives the time-honored leading
non-analytic contribution
\begin{eqnarray}
\label{dim}
 I_N^{\pi}=- \frac{c}{32\pi} \, M_\pi^3~.
\end{eqnarray}
Note that in DR no power-law divergences appear and therefore loop
graphs can not renormalize the baryon mass in the chiral limit and the
dimension two LECs which leads to self-energy term $\sim M_\pi^2$.
If we instead use a three-momentum cut-off as suggested in \cite{BHMcut},
the same diagram leads to the expression 
\begin{eqnarray}
  I_{N}^\pi 
  & = & -\frac{c}{16\pi^2}\int_0^\infty dy \int_0^\Lambda d|k|
  \frac{\vec{k}^4}{(\vec{k}^2+M_\pi^2+y^2)^{3/2}} \nonumber\\
   & = & -\frac{c}{16 \pi^2}\left\{\frac{\Lambda^3}{3}-M_\pi^2 \Lambda
  -M_\pi^3\arctan\frac{M_\pi}{\Lambda} \right\}-\frac{c}{32\pi}M_\pi^3~.
\end{eqnarray}
Note that besides the contribution $\sim M_\pi^3$ that is free of the cut-off,
we have additional divergent and finite terms. The cubic divergence independent
of the pion mass leads to a renormalization of the baryon mass in the
chiral limit whereas the term $\sim M_\pi^2 \Lambda$ renormalizes the dimension
two LECs (the precise relations between the bare and the renormalized
parameters will be given in the following section). Also, the finite 
arctan term is formally of higher order since
it only starts to contribute at order $M_\pi^4$. Chiral symmetry has been
manifestly maintained by this procedure since no structures besides the 
ones already appearing in the effective Lagrangian are needed to absorb all
divergences (see also \cite{BHMcut} and the more systematic work reported in
\cite{MZcut}). Also, the DR result can be formally obtained when letting
the cut-off tend to infinity. This will also be discussed in more detail
below.

\section{Formalism II: Baryon masses}
\def\theequation{\arabic{section}.\arabic{equation}}
\setcounter{equation}{0}
\label{sec:masses}

This section contains the basic formalism to calculate the baryon masses
to fourth order in the chiral expansion. We briefly discuss the third
order result and the introduction of an improvement term as proposed
in \cite{BHMcut}. We then proceed to present the central new results,
namely the baryon masses to fourth order utilizing cut-off regularization.
The calculation of the baryon self-energy and the corresponding
mass shift at a given order in the chiral expansion is briefly outlined
in App.~\ref{app:masses}.

\subsection{Baryon masses at third order}
\label{sec:masses3}

The calculation of the baryon masses to third order in cut-off regularization
is straightforward. Utilizing the loop integrals collected in
App.~\ref{app:int}, one obtains
\begin{eqnarray}
\label{m3}
  m_B & = & m_0^{(r)} + \gamma_B^D b_D^{(r)} + \gamma_B^F b_F^{(r)} 
-2 b_0^{(r)} (M_\pi^2 + 2 M_K^2)
 - \frac{1}{24\pi F_0^2}\left[\alpha_B^\pi M_\pi^3 + \alpha_B^K M_K^3 +
     \alpha_B^\eta M_\eta^3\right]\nonumber\\
    & + &  \frac{1}{12 \pi^2 F_0^2} \bigg[\alpha_B^\pi M_\pi^3 
\arctan\frac{M_\pi}{\Lambda} + \alpha_B^K M_K^3 \arctan\frac{M_K}{\Lambda}
   +  \alpha_B^\eta M_\eta^3 \arctan\frac{M_\eta}{\Lambda}\bigg] + {\cal O}(q^4)~,
\end{eqnarray}
where the state-dependent coefficients $\gamma_B^{D,F}$ and $\alpha_B^P$ can
be found e.g. in \cite{BKMrev}. As announced, the baryon mass and the
the dimension two couplings are renormalized as symbolized by the 
superscript $(r)$. The precise renormalization takes the form 
\begin{eqnarray}
  m_0^{(r,3)} & = & m_0-\left(\frac{5 D^2}{36 \pi^2 F_0^2} 
+\frac{ F^2}{4 \pi^2 F_0^2}\right)\Lambda^3~, \quad
  b_F^{(r,3)}  =  b_F-\left(\frac{5 D F}{48 F_0^2 \pi^2}\right)\Lambda,
\nonumber\\ 
  b_0^{(r,3)} & = & b_0 -\left(\frac{13 D^2 + 9 F^2}{144 \pi^2 F_0^2}
\right)\Lambda~,\quad 
  b_D^{(r,3)}  =  b_D -\left(\frac{-D^2 + 3 F^2}{32 \pi^2 F_0^2}\right)\Lambda.
\end{eqnarray}
It is instructive to expand the $M_P^3 \arctan (M_P/\Lambda)$ 
($P = \{\pi, K,\eta\}$) contributions
\begin{eqnarray}
\label{m3ent}
  m_B & = & m_0^{(r)} + \gamma_B^D b_D^{(r)} + \gamma_B^F b_F^{(r)} 
-2 b_0^{(r)} (M_\pi^2 + 2 M_K^2)
- \frac{1}{24\pi F_0^2}\left[\alpha_B^\pi M_\pi^3 + \alpha_B^K M_K^3 
+ \alpha_B^\eta M_\eta^3\right]\nonumber\\
    & + &  \frac{1}{12 \pi^2 F_0^2} \alpha_B^\pi M_\pi^3 
\left\{ \frac{M_\pi}{\Lambda}-\frac{1}{3}\left(\frac{M_\pi}{\Lambda}\right)^3
+\ldots \right\}
  + \frac{1}{12 \pi^2 F_0^2} \alpha_B^K M_K^3 \left\{\frac{M_K}{\Lambda}
-\frac{1}{3}\left(\frac{M_K}{\Lambda}\right)^3+\ldots\right\}\nonumber\\
  & + & \frac{1}{12 \pi^2 F_0^2} \alpha_B^\eta M_\eta^3 
\left\{\frac{M_\eta}{\Lambda}
-\frac{1}{3}\left(\frac{M_\eta}{\Lambda}\right)^3+\ldots\right\}\nonumber\\
   & = & m_0^{(r)} + \gamma_B^D b_D^{(r)} + \gamma_B^F b_F^{(r)} 
-2 b_0^{(r)} (M_\pi^2 + 2 M_K^2)
- \frac{1}{24\pi F_0^2}\left[\alpha_B^\pi M_\pi^3 + \alpha_B^K M_K^3 
+ \alpha_B^\eta M_\eta^3\right]+O(q^4)~,
\end{eqnarray}
where the last line corresponds to the DR result. As stated earlier, the
additional contributions scale with inverse powers of the cut-off and
thus vanish when $\Lambda \to \infty$ (which formally corresponds to DR).
As already stressed in \cite{BHMcut} (and noted by others), this third
order representation is not sufficiently accurate to make connection
with lattice results if one is not very close to the physical value of the
GB masses. Therefore, one should perform a fourth order calculation or at 
least add an improvement term that is formally of fourth order but should
be elevated to third order. We first discuss briefly this latter possibility
before turning to the full-fledged fourth order calculation.

\subsection{Improvement term}
\label{sec:mass3imp}

\begin{figure}[htb]
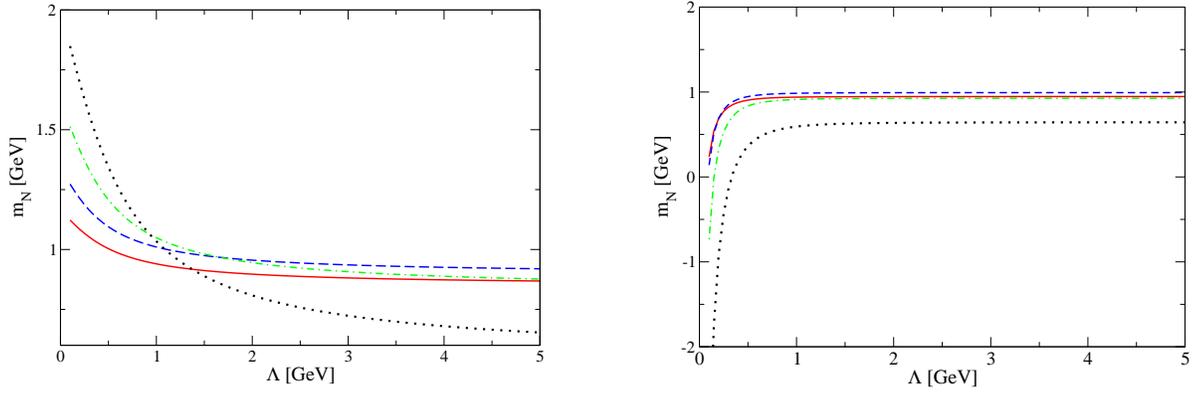

\parbox{.50\textwidth}{\epsfig{file= Ncut3.eps,width=.42\textwidth,angle=0,
silent=,clip=}}
\hfill
\parbox{.50\textwidth}{\epsfig{file= Ncut3i.eps,width=.42\textwidth,angle=0,
silent=,clip=}}
\vspace{0.2cm}
\begin{center}
\caption{Cut-off dependence of the nucleon mass for various pion masses
with the kaon mass fixed. Solid/dashed/dot-dashed/dotted line: $M_\pi =
140/300/450/600\,$MeV. Left panel: Third order calculation. Right panel:
Third order calculation with the improvement term.
\label{fig:platNpi3}}
\end{center}
\end{figure}

As noticed in \cite{BHMcut}, the third order result for the baryon masses
shows a very strong cut-off dependence when the pion mass is increased above
its physical value, see the left panel in Fig.~\ref{fig:platNpi3}. Only when
one has a plateau below the chiral symmetry breaking scale $\Lambda_\chi
= 4\pi F_\pi \simeq 1.2\,$GeV, one has the required cut-off independence.
For pion masses above 300~MeV, this plateau vanishes. 
To obtain a better stability against cut-off variations, 
it was therefore proposed in  \cite{BHMcut} to promote the fourth order 
operator $e_1 M_\pi^4 \bar N N$ to the third order and to cancel the leading
cut-off dependence in Eq.~(\ref{m3ent}) by a proper adjustment of the LEC
$e_1$, $e_1 = e_1^{\rm fin} - {\rm coeff} / \Lambda$, where the
coefficient can be read off from the leading term of the expansion of the 
arctan function. For the SU(3) calculation performed here, the situation
is a bit more complicated. In fact, the corresponding improvement term
for the baryon $B$ consists of three contributions (we use the notation
of \cite{BM})
\begin{eqnarray}\label{imp3}
\epsilon_{1,B}^{\pi\pi} M_\pi^4 +\epsilon_{1,B}^{\pi K} M_\pi^2 M_K^2 
+\epsilon_{1,B}^{KK} M_K^4~,
\end{eqnarray}
where the coefficients $\epsilon_{1,B}^{PQ}$ ($P,Q = \{\pi,K\}$)
are linear combinations
of the fourth order LECs $d_i$ defined in Eq.~(\ref{L4}) (the precise
relation can be found again in \cite{BM}). Throughout, we use the
Gell-Mann--Okubo relation to express the $M_\eta$-term by the pion and the
kaon masses, $3M_\eta^2 = 4M_K^2 -M_\pi^2$. To eliminate the leading 
$1/\Lambda$ dependences, the LECs $d_i$ have to take the form
\begin{eqnarray}\label{diL}
 d_1 & = & d_1^{\rm fin} -\frac{D^2 - 3 F^2}{576  \pi^2 F_0^2 \Lambda}~,\quad
 d_2   =   d_2^{\rm fin} -\frac{D F}{64  \pi^2F_0^2 \Lambda}~,\quad
 d_3   =   d_3^{\rm fin} -\frac{D^2 - 3 F^2}{128 \pi^2 F_0^2 \Lambda}~,\quad
 d_4   =   d_4^{\rm fin} -\frac{-D^2 + 3 F^2}{64 \pi^2 F_0^2 \Lambda}~,\nonumber\\
 d_5  & = & d_5^{\rm fin}+ \frac{13 D F}{288 \pi^2 F_0^2 \Lambda}~,\quad
 d_7  =  d_7^{\rm fin}+ \frac{35 D^2 + 27 F^2}{3456 \pi^2 F_0^2 \Lambda}~,\quad
 d_8  =  d_8^{\rm fin}+\frac{17 D^2 + 9 F^2}{1152 \pi^2 F_0^2 \Lambda}~.
\end{eqnarray}

\subsection{Baryon masses at fourth order}
\label{sec:masses4}

To fourth order in the chiral expansion, the octet baryon masses can be
written as (when employing CR)
\begin{eqnarray}
 m_B & = & m_0^{(r)} + \delta m_B^{(2)} + \delta m_B^{(3)} + \delta m_B^{(4)}\nonumber\\
     & = & m_0^{(r)} + \delta m_B^{(2)} +\Delta m_B^{(3)}+f_{B,3}(\Lambda)
+\Delta m_B^{(4)} + f_{B,4}(\Lambda)~,
\end{eqnarray}
where in the second line we have split the mass shift $\delta m_B^{(i)}$ ($i =
3,4$) into cut-off independent  and an explicitly cut-off dependent piece. This
is done to facilitate the comparison with the results obtained in DR.
The various pieces take the form
\begin{eqnarray}
\label{masseges}
  \delta m_B^{(2)} & = & \gamma_B^D b_D^{(r)} + \gamma_B^F b_F^{(r)} 
-2 b_0^{(r)} (M_{0,\pi}^2 + 2 M_{0,K}^2)~,
\nonumber\\
   \Delta m_B^{(3)}  & = & -\frac{1}{24\pi F_0^2}\bigg[\alpha_B^\pi M_\pi^3 + \alpha_B^K M_K^3 
                         + \alpha_B^\eta M_\eta^3\bigg],\nonumber\\
    f_{B,3}(\Lambda)  & = &  \frac{1}{12 \pi^2 F_0^2} \bigg[\alpha_B^\pi
    M_\pi^3 \arctan\frac{M_\pi}{\Lambda} 
                       + \alpha_B^K M_K^3 \arctan\frac{M_K}{\Lambda}
                   +  \alpha_B^\eta M_\eta^3 \arctan\frac{M_\eta}{\Lambda}\bigg]~,\nonumber\\
    \Delta m_B^{(4)}  & = & \epsilon_{1,B}^{P,Q} M_P^2 M_Q^2 
                         + \epsilon^{P,Q}_{2,B,T} M_P^2 M_Q^2 \ln\frac{M_T}{m_0}~,\nonumber\\
  f_{B,4}(\Lambda)  
               & = & -\epsilon_{2,B,T}^{P,Q} M_P^2 M_Q^2 \ln\left(1+R_T\right)
                 +  \frac{1}{R_T}
                      \Bigg\{\beta_{1,B,T} \Lambda^4\left( 1 - R_T
                   +   \frac{1}{2}\left(\frac{M_T}{\Lambda}\right)^2\right)\nonumber\\
                 &&  +   \beta_{2,B,T}^P M_P^2 \Lambda^2\left(1-R_T\right)
                  +  \beta_{3,B,T}^{P,Q} M_P^2 M_Q^2\Bigg\}~,
\end{eqnarray}
with $R_T = (1 + M_T^2/\Lambda^2)^{1/2}$.
Here, we have $P,Q,T = \{ \pi, K,\eta\}$ and the summation convention for
these indices is understood. To arrive at these results, we made use of the
loop integrals collected in App.~\ref{app:int}. The resulting
$\Lambda$--dependent terms are separated into a contribution that contains all
the power and logarithmic divergences and another one that is finite in the
limit that  $\Lambda\to \infty$. Only these latter terms are displayed
here. For the nucleon, the $\epsilon$--coefficients  and the
$\beta$--coefficients are collected in App.~\ref{app:masses}. Note also that the fourth
order LECs $d_i$ have the form displayed in Eq.~(\ref{diL}) so that the 
leading cut-off dependence is canceled. In the second order term, we have
used the leading terms in the quark mass expansion of the pion and the kaon
mass, denoted   $M_{0,P}$, because the difference to the physical masses
appears in the fourth order of the baryon mass expansions. For consistency,
we have recalculated the Goldstone boson masses in cut-off regularization
(this was not done e.g. in \cite{BHMcut}), the explicit formulae are given
in App.~\ref{app:mesonmasses}. All appearing polynomial and logarithmic
divergences are absorbed in a consistent redefinition of the bare parameters.
The renormalization of the chiral limit mass and of the LECs takes the
form
\begin{eqnarray}
\label{ren4}
   m_0^{(r,4)}  &=&  m_0^{(r,3)} + \frac{m_0}{\pi^2 F_0^2}
               \bigg(-\frac{b_{12}}{8}-\frac{3 b_4}{4} - \frac{7 b_7}{12} 
               - b_9 + \frac{3 b_{15} m_0}{4}+ \frac{7 b_{17} m_0}{12}
               +\frac{b_{18} m_0}{8}
              + b_{19} m_0 \bigg) \Lambda^4,\nonumber\\
   b_F^{(r,4)}  &=& b_F^{(r,3)}-\frac{1}{96 \pi^2 F_0^2}
                   \Big(-10 b_2 + 14 b_F + 10 b_F D^2 +
                    20 b_D D F 
                    + 18 b_F F^2\nonumber\\
                &&   -5 b_5 m_0  - 5 b_6 m_0 
                    + 5 b_{16} m_0^2 \Big)\Lambda^2,\nonumber\\
   b_0^{(r,4)} &=&  b_0^{(r,3)} -\frac{1}{144 \pi^2 F_0^2}
                      \Big(- 18 b_1 
                       - 26 b_3  
                       - 48 b_8  + 18 b_D  + 
                        48 b_0  - 26 b_D D^2 
                       - 36 b_F D F
                       - 18 b_D F^2 \nonumber\\ 
           &&          - 9 b_4 m_0 - 13 b_7 m_0 - 
                       24 b_9 m_0 + 9 b_{15} m_0^2
                     + 13 b_{17} m_0^2 + 
                      24 b_{19} m_0^2\Big)\Lambda^2,\nonumber\\
   b_D^{(r,4)} &=& b_D^{(r,3)}-\frac{1}{96 \pi^2 F_0^2}
                       \Big( - 18 b_1  - 2 b_3  
                       + 14 b_D 
                       + 26 b_D D^2  + 
                       36 b_F D F \nonumber\\
            &&         + 18 b_D F^2  - 3 b_{12} m_0 - 
                       9 b_4 m_0
                       -  b_7 m_0 + 9 b_{15} m_0^2 +  b_{17} m_0^2 + 
                       3 b_{18} m_0^2\Big)\Lambda^2,\nonumber\\
   d_1^{(r,4)}&=&d_1-\frac{1}{1152 \pi^2 F_0^2 m_0}
                  \Big(-D^2 + 3 F^2 + 12 b_1 m_0 - 4 b_3 m_0 
                   - 14 b_D m_0\nonumber\\ 
               &&      - 69 b_D D^2 m_0 - 162 b_F D F m_0 - 81 b_D F^2 m_0 + 
                       3 b_4 m_0^2 - b_7 m_0^2 - 3 b_{15} m_0^3
                     + b_{17} m_0^3\Big)\ln\frac{\Lambda}{m_0},\nonumber\\ 
   d_2^{(r,4)}&=& d_2-\frac{1}{768 \pi^2 F_0^2 m_0}\Big(-6 D F - 
              12 b_2 m_0 - 4 b_F m_0 - 60 b_F D^2 m_0 \nonumber\\
             && - 120 b_D D F m_0 - 108 b_F F^2 m_0 - 3 b_5 m_0^2 
                - 3 b_6 m_0^2 + 
              3 b_{16} m_0^3\Big)\ln\frac{\Lambda}{m_0},\nonumber\\
   d_3^{(r,4)}&=& d_3-\frac{1}{768 \pi^2 F_0^2 m_0}(-3 D^2 + 
              9 F^2 + 36 b_1 m_0 + 4 b_3 m_0 - 24 b_D m_0
              - 78 b_D D^2 m_0 \nonumber\\
             && - 108 b_F D F m_0 - 54 b_D F^2 m_0 + 
              3 b_{12} m_0^2 + 9 b_4 m_0^2 + b_7 m_0^2
              - 9 b_{15}{m_0^3} - b_{17} m_0^3 - 3 b_{18} m_0^3\Big)
               \ln\frac{\Lambda}{m_0},\nonumber\\
   d_4^{(r,4)}&=& d_4 -\frac{1}{1152 \pi^2 F_0^2 m_0}\Big(9 D^2 - 
              27 F^2 - 108 b_1  m_0+ 4  b_3 m_0 + 44 b_D m_0\nonumber\\
             && + 288 b_D D^2 m_0 - 6 b_{12} m_0^2 - 27 b_4 m_0^2 
                + b_7 m_0^2 + 
              27 b_{15} m_0^3 - b_{17} m_0^3
            + 6 b_{18} m_0^3\Big)\ln\frac{\Lambda}{m_0},\nonumber\\
   d_5^{(r,4)}&=& d_5 -\frac{1}{1152 \pi^2 F_0^2 m_0}\Big(26 D F + 
              52 b_2 m_0 - 44 b_F m_0 + 13 b_5 m_0^2 + 13 b_6 m_0^2
            - 13 b_{16} m_0^3\Big)\ln\frac{\Lambda}{m_0},\nonumber\\
   d_7^{(r,4)}&=& d_7-\frac{1}{6912 \pi^2 F_0^2 m_0}\Big(35 D^2 + 
              27 F^2 + 108 b_1 m_0 + 140 b_3 m_0 + 264 b_8 m_0\nonumber\\
               && - 
              132 b_D m_0 - 264 b_0 m_0 + 144 b_D D^2 m_0  + 27  b_4 m_0^2 + 
              35 b_7 m_0^2 + 66 b_9 m_0^2\nonumber\\
        && - 27 b_{15} m_0^3 - 
              35 b_{17} m_0^3 - 66 b_{19} m_0^3\Big)
              \ln\frac{\Lambda}{m_0},\nonumber\\
    d_8^{(r,4)}&=&d_8-\frac{1}{2304 \pi^2 F_0^2 m_0}\Big(17 D^2 + 
              9 F^2 + 36 b_1 m_0 + 68 b_3 m_0 + 120 b_8 m_0\nonumber\\
            && - 28 b_D m_0 - 
              120 b_0 m_0 + 168 b_D D^2 m_0 + 432 b_F D F m_0 + 
              216 b_D F^2 m_0\nonumber\\
              && + 9 b_4 m_0^2 + 17 b_7 m_0^2 + 
              30 b_9 m_0^2 - 9 b_{15} m_0^3 - 17 b_{17} m_0^3 - 
              30 b_{19} m_0^3\Big)\ln\frac{\Lambda}{m_0}~.
\end{eqnarray}
Clearly, higher order powers in the cut-off $\Lambda$ appear as compared to
the third order calculation. Note also the appearance of a new scale in the
logarithmically divergent terms. As can be seen from App.~\ref{app:int}, the
integrals $I_1 - I_6$ and $\alpha_1 - \alpha_3$ contain terms proportional to
$\ln (M_P/\Lambda)$. To properly absorbs these divergences, one uses
$\ln (M_P/\Lambda) = \ln(M_P/\nu) - \ln(\Lambda/\nu)$, with $\nu$ the new
scale. From here on, we set $\nu = m_0$. As a check, it can be shown that
for $\Lambda \to \infty$ our fourth order results agree with the ones
obtained in DR in \cite{BM} if one sets $\mu = m_0$, with $\mu$ the scale of DR.
For this comparison to work, one also has to account for the fact that in \cite{BM}
some terms $\sim M_P^2 M_Q^2$ were absorbed in a redefinition of the $d_i$.
This concludes the necessary formalism, we now
turn to the numerical analysis.

\section{Results and discussion}
\def\theequation{\arabic{section}.\arabic{equation}}
\setcounter{equation}{0}
\label{sec:res}

Before presenting results, we must fix parameters.
In the meson sector, we use standard values of the LECs 
$L_i$ at the scale $M_\rho$: $L_4 = -0.3, L_5 = 1.4, L_6 = 
-0.2, L_7 = -0.4$ and $L_8 = 0.9$ (all in units of $10^{-3}$).
These are run to the scale $\lambda = m_0$ using the standard
one-loop $\beta$--functions. Throughout, we set $F_0 = 100\,$ MeV,
which is an average value of the physical values of $F_\pi, F_K$ 
and $F_\eta$. For the leading baryon axial couplings we use
$ D=0.75$ and $F=0.5$. We have checked that varying these parameters within 
phenomenological bounds does not alter our conclusions.

\subsection{Fixing the low-energy constants} 
\label{sec:LECs}

For the dimension two LECs from the meson-baryon Lagrangian
we use the central values of \cite{BM}, these are collected
in Tab.~\ref{LECb}. We have not varied these LECs since such
modifications can effectively be done by changing the fourth order
LECs $d_i$ within reasonable bounds.
\begin{table}[htb]
\begin{center}
  \begin{tabular}{|c|c|c|c|c|c|c|}\hline
$b_0$ & $b_D$ & $b_F$ & $b_1$ & $b_2$ & $b_3$ & $b_8$ \\ \hline
 -0.606     & 0.079 & -0.316 & -0.004 & -0.187 & 0.018 & -0.109  \\  \hline
\end{tabular}
\caption{\label{LECb} Values of the LECs $b_i$ in $\mbox{GeV}^{-1}$ taken 
from \protect\cite{BM}.}
\vspace{-0.2cm}
\end{center}
\end{table}
Next we consider the determination of the fourth order LECs $d_i$. As noted
before, the improvement term for each baryon consists of three pieces. To
determine these, we have only considered  the nucleon, for two reason. First,
there are more lattice results for this particle than for the others and
second, it also facilitates the direct comparison with the SU(2) results of
\cite{BHMcut}. For the nucleon, the coefficients appearing in Eq.~(\ref{imp3})
are related to the LECs $d_i$ via
\beq
\epsilon_{1,N}^{\pi\pi} = -4 (4 d_1 + 2 d_5 + d_7 + 3 d_8 )~,\quad
\epsilon_{1,N}^{\pi K}  = 8 (4 d_1 -2 d_2 - d_5 - 2 d_7 + 2 d_8)~,\quad
\epsilon_{1,N}^{KK}     =  -16 ( d_1 - d_2 + d_3 - d_5 + d_7 + d_8)~.
\eeq
We have now varied the values of the $d_i$ under the following constraints:
We require that $m_N (M_\pi)$ passes through the physical value $m_N =
940\,$MeV for the physical pion mass $M_\pi =140\,$MeV and similarly for
$m_N (M_K)$ at $M_K = 494\,$MeV. Furthermore, we only allow for variations 
of  $\delta d_i = \pm 0.1\,$GeV$^{-3}$ from the central values of \cite{BM}
(this is in fact the largest magnitude of any of these LECs). Under
these restrictions, we have tried to describe the trend of the earlier 
MILC data with the third order improved formula (note that the more
recent data \cite{MILCnew} were not used in the fit for reasons discussed
below). Note that we have reconstructed these data from table~IX~(VII) of 
Ref.~\cite{MILC}(\cite{MILCnew}) using the scale parameter $r_1 = 0.35
(0.317)\,$fm. The resulting set of values for the
LECs $d_i$ is denoted as the ``optimal set'' from here on. In Tab.~\ref{LECd}
we have collected the values of the $d_i$ from \cite{BM} and for the optimal
set. In fact, the values of the $d_i$ given in that table refer to the basis
used in \cite{BM} as indicated by the superscript ``BM''. These differ by some
small finite shifts from the one used in CR (we refrain from giving the explicit 
formulae here).   Note that one can not exactly reproduce these
lattice data as shown by the solid line in Fig.~\ref{fig:mN3imp}, but the
nucleon mass now increases with growing pion mass as demanded by all existing
lattice results. Also, at fourth order there are other contributions which are
not captured by the improvement term, see the discussion below.  We also note
that we have not restricted the $d_i$ such that the GMO relation $3m_\Lambda + m_\Sigma
= 2(m_N+m_\Xi)$ is fulfilled 
so as to have a better handle on the theoretical uncertainty. Including the
improvement term with these values of the LECs leads indeed to a very reduced
cut-off dependence as shown in the right panel of Fig.~\ref{fig:platNpi3}.
Note that the treatment of the improvement term is more tricky in SU(3) than
in the SU(2) calculation since one has to balance three different terms as 
opposed to fixing one in the two-flavor case.
\begin{table}[htb]
\begin{center}
  \begin{tabular}{|c|c|c|c|c|c|c|c|}\hline
   &$d_1^{\rm BM}$ & $d_2^{\rm BM}$ & $d_3^{\rm BM}$ 
   & $d_4^{\rm BM}$ & $d_5^{\rm BM}$ & $d_7^{\rm BM}$ & $d_8^{\rm BM}$ \\ \hline
  BM \protect\cite{BM}
       & $\phantom{-}$0.008 & $\phantom{-}$0.035 &$\phantom{-}$0.069 &
   $-$0.077 &  $-$0.05 & $-$0.018  &
  $-$0.103 \\   \hline
  opt. & $-$0.043 & $-$0.066 &   $-$0.031  & $-$0.077  &  $-$0.15 & $-$0.118  
  & $-$0.2\\ \hline
\end{tabular}
\caption{\label{LECd} Values of the LECs $d_i$ in $\mbox{GeV}^{-3}$ in the
basis used in \protect\cite{BM} as indicated by the superscript ``BM''. ``Opt.''
denotes the optimal set as described in the text.}
\end{center}
\end{table}

\subsection{Nucleon mass and pion-nucleon sigma term} 
\label{sec:mN}

\begin{figure}[H]
\vspace{0.04cm}
\centerline{
\epsfig{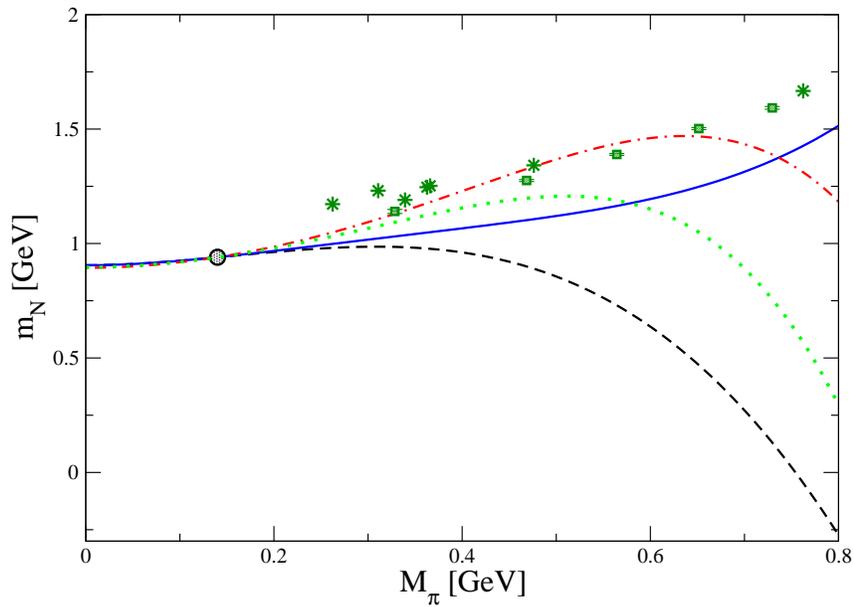}
}
\vspace{0.2cm}
\begin{center}
\caption{Nucleon mass in DR at third (dashed), improved third
  (solid) and fourth (dot-dashed) order, respectively. The dotted line
  represents the fourth order calculation from \protect\cite{BM}.
The three flavor data are from the MILC collaboration (boxes from
\protect\cite{MILC} and stars from \protect\cite{MILCnew}).
The filled circle gives the value of the physical nucleon mass at
the physical value of $M_\pi$.
\label{fig:mN3imp}}
\end{center}
\end{figure}

\begin{figure}[H]
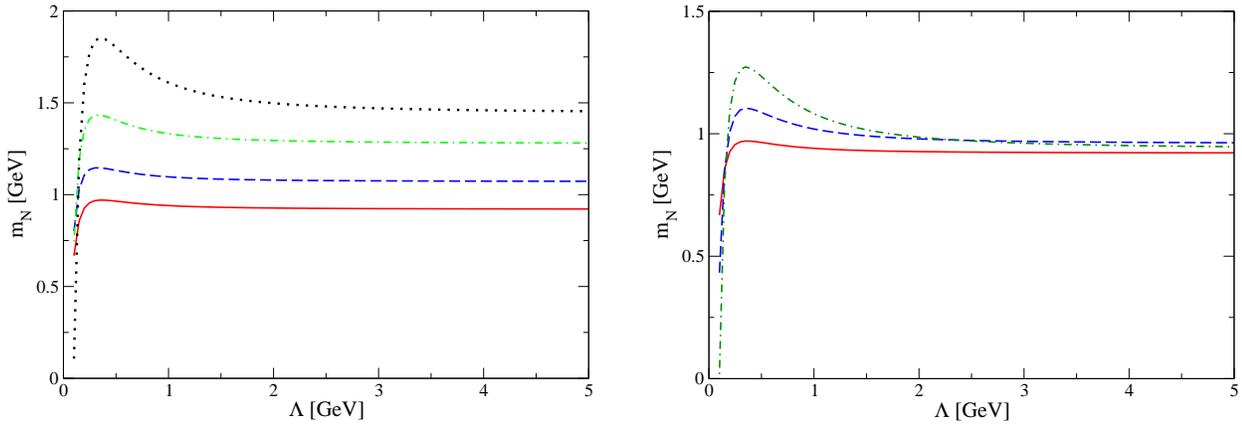

\vspace{0.9cm}
\parbox{.50\textwidth}{\epsfig{file= Ncut4.eps,width=.46\textwidth,angle=0,
silent=,clip=}}
\hfill
\parbox{.50\textwidth}{\epsfig{file= Ncut4k.eps,width=.46\textwidth,angle=0,
silent=,clip=}}
\vspace{0.2cm}
\begin{center}
\caption{Left panel: Cut-off dependence of the nucleon mass for various pion masses
with the kaon mass fixed. Solid/dashed/dot-dashed/dotted line: $M_\pi =
140/300/450/600\,$MeV. Right panel: Kaon mass dependence for fixed pion mass.
Solid/dashed/dot-dashed line: $M_K = 494/600/700$~MeV.
\label{fig:platNpi4}}
\end{center}
\end{figure}

We now consider the nucleon mass as function of the pion and the kaon
mass in DR and CR. We have calculated $m_N$ at third, improved third
and fourth order, see Fig.~\ref{fig:mN3imp} for DR. In what follows,
we mostly focus on the results obtained at fourth order. The trends
when going from third to improved third to fourth order for the pion mass
dependence of the nucleon mass are very similar to the SU(2) case discussed
in big detail in \cite{BHMcut}. We see that with the optimal set of
the $d_i$ as given in Tab.~\ref{LECd} one obtains a rather accurate
description of the earlier and most of the more recent MILC data
\cite{MILCnew} for 
pion masses below 600~MeV, cf. the dot-dashed line in Fig.~\ref{fig:mN3imp}. 
Note, however, that the two lowest pion mass points of the recent MILC
data \cite{MILCnew} do not quite fit into the trend of our extrapolation function if one
insists that for the physical pion mass the curve runs through the physical 
value of $m_N$. More low mass pion data and/or a more sophisticated treatment 
of finite size/volume effects are needed to resolve this
problem. 
If one were to use the
$d_i$ determined in \cite{BM}, one already deviates sizeably from the
trend of the MILC data for pion masses starting at about 500~MeV (dotted line
in Fig.~\ref{fig:mN3imp}). The same can be seen for the fourth order calculation based
on CR utilizing $\Lambda = 1\,$GeV in Fig.~\ref{fig:mN4} (left panel). To get
a better idea about the uncertainty when going to higher pion masses, we
have also performed calculations with three other sets, namely setting all
$d_i = 0.2/0/-0.2\,$GeV$^{-3}$, corresponding to the long-/medium-/short-dashed
lines in that figure. This clearly overestimates the theoretical uncertainty
since some of the $d_i$ are correlated parameters. Still it is safe to say that
for pion masses below 550~MeV the theoretical error is moderately small.
These results for the pion mass dependence of $m_N$ as well as for its
cut-off dependence at a given pion mass are very similar to the results
of two-flavor study reported in  \cite{BHMcut}.
In the right panel of Fig.~\ref{fig:mN4}, we show the kaon mass dependence for
the same variety of choices for the $d_i$. Since we enforce that $m_N$ takes
its physical value for $M_K =494\,$MeV, the resulting kaon mass dependence is
much flatter than the pion mass dependence with decreasing meson masses.   
In the left (right) panel of Fig.~\ref{fig:platNpi4} we show the cutoff dependence
of $m_N$ for various values of the pion (kaon) mass. For pion masses up to 450~MeV,
one has a nice plateau below the scale of chiral symmetry breaking but not any
more for $M_\pi = 600$~MeV. For the kaon mass dependence, the situation is somewhat
different due to the much larger meson mass. Here, we still have a reasonable plateau
at $M_K \simeq 600\,$MeV. These observations are consistent with our earlier
observations that chiral extrapolations in $M_\pi$ based on the fourth order 
CHPT representation can be applied for masses up to 550~MeV, a result which is 
consistent with the one found for the SU(2) calculation in \cite{BHMcut}. 
For the kaon mass dependence, one can even go to somewhat higher meson masses.

\medskip
\noindent
It is also interesting to study the range of values found for the octet chiral
limit mass $m_0$ at the various orders and employing the different regularization 
schemes and values of the LECs $d_i$. One observes that $m_0$ increases with
increasing cut-off, that means in DR its value is above the one in CR when
one chooses $\Lambda = 1\,$GeV. Insisting that $m_N$ takes its physical value
at the physical value of $M_\pi$ and $M_K$ when studying the pion and the kaon
mass dependence, respectively, we find
\beq\label{rangem0}
710~{\rm MeV} \lesssim m_0 \lesssim 1070~{\rm MeV}~,
\eeq  
which is consistent with expectations and also with the findings in 
\cite{BM} (note that in that paper a different method was used to estimate
the theoretical uncertainty, which we consider less reliable than the
one used here). The range given in Eq.~(\ref{rangem0}) of course includes
the SU(2) value of about 880~MeV \cite{BHMcut}. Also, we note again that
the MILC data are obtained using staggered fermions, so strictly speaking one
should use ``staggered fermion CHPT''. However, we believe that
this will not significantly alter the trends discussed here.

\medskip
\noindent
Another quantity of interest is the pion-nucleon sigma term,
\beq
\sigma_{\pi N} (0) = \hat{m} \langle N |\bar uu + \bar dd|N\rangle
= \hat{m} \frac{\partial m_N}{\partial \hat m} 
= M_\pi^2  \frac{\partial m_N}{\partial M_\pi^2}~.
\eeq
It can be extracted directly from the the slope of $m_N (M_\pi)$ at $M_\pi =
0$. For the optimal set and the LECs from \cite{BM}, we obtain the range 
(considering DR and CR with $\Lambda = 1\,$GeV)
\beq
\sigma_{\pi N} (0) = 50.7 \ldots 53.7~\mbox{MeV}~.
\eeq
These numbers are consistent with the results of \cite{BM} (as they should) and
the study of SU(2) lattice data in \cite{PHM}, $\sigma_{\pi N} = 49\pm 3\,$MeV.
The resulting strangeness fraction can be obtained from $\sigma_{\pi N} (0) =
\sigma_0 /(1-y)$ and using $\sigma_0 = 37\,$MeV from \cite{BM} (which is
consistent with the pioneering work in \cite{JG}, $\sigma_0 = 35\pm 5\,$MeV).
This leads to $y = 0.27 \ldots  0.31$, which is again consistent with
\cite{BM} but somewhat on the large side.

\subsection{Hyperon  masses} 
\label{sec:mB}

We now consider the octet members with strangeness. As noted before, when fixing
the coefficients in the improvement term, we have not insisted to recover the 
Gell-Mann--Okubo relation, thus some of the masses are somewhat off their
empirical values. In Tab.~\ref{massevgl} we collect the resulting values for
the improved third and fourth order. While the $\Sigma$ mass is well
reproduced, the $\Lambda$ and $\Xi$ masses come out by about $10 - 15$\% too
high. To get a  handle on the theoretical accuracy, we also use the values
for the $d_i$ from \cite{BM}, in that case all masses are exactly reproduced.
\begin{table}[htb]
\begin{center}
\begin{tabular}{|c|c|c|c|c|}\hline
order /   & imp. 3rd & fourth  & fourth & exp.\\ 
baryon    & CR   & CR   & DR   & \\ \hline
$\Lambda$ & 1115 & 1304 & 1243 & 1116\\ \hline
$\Sigma$  & 1101 & 1194 & 1167 & 1193\\ \hline
$\Xi$     & 1222 & 1532 & 1437 & 1315\\ \hline
\end{tabular}
\caption{\label{massevgl} Baryon masses in MeV in DR and CR with
$\Lambda = 1\,$GeV for 
different orders. For the experimental numbers, we haven taken the
masses of the neutral particles.}
\end{center}
\end{table}

\noindent The corresponding pion and kaon mass dependences for the $\Lambda$,
the $\Sigma$ and the $\Xi$ are shown in Figs.~\ref{fig:hypL},~\ref{fig:hypS}
and \ref{fig:hypX}, respectively. The solid/dashed lines refer to the optimal
set of the LECs/to the LECs from \cite{BM}. We note in particular that the pion mass
dependence for the $\Xi$ is much flatter as one would expect from the MILC
data. This is not unexpected -- the $\Xi$ only contains one valence light quark
and should thus be less sensitive to variations in the pion mass. Clearly, one
could improve this description by fitting directly to these particles.

\bigskip

\section*{Acknowledgements}

We are grateful to Claude Bernard for some clarifying comments on the first
version of this manuscript and Thomas Lippert for a very useful communication.

\vfill

\pagebreak

\appendix
\def\theequation{\Alph{section}.\arabic{equation}}
\setcounter{equation}{0}
\section{Baryon masses}
\label{app:masses}

Here, we collect some formalism to calculate the baryon masses from the baryon
self-energies. Consider the heavy baryon approach. The baryon four-momentum
is $p_\mu = m_0 v_\mu + r_\mu$, with $v_\mu$ the four-velocity subject to the
constraint $v^2 =1$ and $r_\mu$ is the (small) residual four-momentum,
$\omega = v \cdot r \ll m_0$.
The baryon self-energy $\Sigma(\omega,r)$ has the chiral expansion
\begin{equation}
\Sigma(\omega,r) = \Sigma^{(2)} (\omega,r)+ \Sigma^{(3)} (\omega,r) +
\Sigma^{(4)} (\omega,r) + \ldots~,
\end{equation}
and the corresponding baryon mass shift is then given by (with $\delta m_B =
\delta m_B^{(2)} + \delta m_B^{(3)} + \delta m_B^{(4)} + \ldots$):
\begin{eqnarray}
  \delta m_B^{(2)} & = & \Sigma^{(2)}(\omega=0,r)+\frac{r^2}{2 m_0},\\
  \delta m_B^{(3)} & = &
  \Sigma^{(3)}(\omega=0,r)+\frac{\partial}{\partial\omega}
  \Sigma^{(2)}(\omega=0,r)\bigg(\delta m_B^{(2)}-\frac{r^2}{2 m_0}\bigg),\\
  \delta m_B^{(4)} &= & -\frac{(\delta m_B^{(2)})^2}{2 m_0}
                    +\Sigma^{(4)}(\omega=0,r)
        +\frac{\partial}{\partial\omega}\Sigma^{(3)}(\omega=0,r) 
                \bigg(\delta m_B^{(2)}-\frac{r^2}{2 m_0}\bigg)\nonumber\\
   &  & +\frac{\partial}{\partial\omega}\Sigma^{(2)}(\omega=0,r)\,
                                                \delta m_B^{(3)}
         +\frac{1}{2}\frac{\partial^2}{\partial\omega^2}
         \Sigma^{(2)}(\omega=0,r)\bigg((\delta m_B^{(2)})^2
          -\delta m_B^{(2)}\frac{r^2}{m_0}+\frac{r^4}{4 m_0^2}\bigg). 
\end{eqnarray}
From this, one obtains the pertinent representations of the baryon masses.
In the following, we only discuss the nucleon mass.
More precisely,  we now give the corresponding non-vanishing prefactors for the 
nucleon (we refrain from  giving the coefficients of the other octet
members).  At third order, cf. Eqs.~(\ref{m3},\ref{m3ent}), one has the 
standard values
\begin{eqnarray}
   \gamma_N^D & = & - 4 M_K^2~, \quad \gamma_N^F  =  4 M_K^2 - 4 M_\pi^2,\nonumber\\
   \alpha_N^\pi & = & \frac{9}{4} (D + F)^2~,\quad
   \alpha_N^K  =  \frac{1}{2} (5 D^2- 6 DF+9F^2)~,\quad
   \alpha_N^\eta  =  \frac{1}{4} (D - 3F)^2~.
\end{eqnarray}
At fourth order, see Eq.~(\ref{masseges}), we have a cut-off independent and a
cut-off dependent contribution to the nucleon mass shift. The coefficients
of the $\Lambda$-independent term read
\begin{eqnarray}
   \epsilon_{1,N}^{\pi\pi} & = &-16 d_1 - 8 d_5 - 4 d_7 - 12 d_8,\nonumber\\
   \epsilon_{1,N}^{\pi K} & = &32 d_1 - 16 d_2 - 8 d_5 - 16 d_7 + 16 d_8,\nonumber\\
   \epsilon_{1,N}^{KK} & = & -16 d_1 + 16 d_2 - 16 d_3 + 16 d_5 - 16 d_7 - 16 d_8,\nonumber\\
   \epsilon_{2,N,\pi}^{\pi\pi} & = &\frac{1}{2(4\pi F_0)^2}\bigg(-12 b_1 - 12 b_2 -12 b_3 - 24 b_8 + 12 b_D + 12 b_F 
                                                                 + 24 b_0 - \frac{3 D^2}{m_0} - \frac{6 D F}{m_0}\nonumber\\ 
        & &\qquad \qquad \quad                                   - \frac{3 F^2}{m_0} -   3 b_4 m_0 - 3 b_5 m_0 
                                                               - 3 b_6 m_0 - 3 b_7 m_0 - 6 b_9 m_0 + 3 b_{15} m_0^2\nonumber\\ 
       & &\qquad \qquad \quad                             + 3 b_{16} m_0^2 + 3 b_{17} m_0^2 + 6 b_{19} m_0^2\bigg), \nonumber\\
%
%
   \epsilon_{2,N,K}^{\pi K} & = & \frac{1}{3(4\pi F_0)^2}\bigg(-52 b_D D^2 + 60 b_F D^2 + 120 b_D D F - 72 b_F D F 
                                                               - 36 b_D F^2 + 108 b_F F^2\bigg),\nonumber\\
    \epsilon_{2,N,K}^{K K} & = & \frac{1}{3(4\pi F_0)^2}\bigg(-36 b_1 + 12 b_2 - 36 b_3 - 48 b_8 + 36 b_D - 12 b_F 
                                                              + 48 b_0 + 52 b_D D^2\nonumber\\
       & &\qquad \qquad \quad                                 -  60 b_F D^2 - 120 b_D D F + 72 b_F D F + 36 b_D F^2 
                                                              - 108 b_F F^2 - \frac{5 D^2}{m_0} \nonumber\\
       & &\qquad \qquad \quad                                 + \frac{6 D F}{m_0} - \frac{9 F^2}{m_0} -  3 b_{12} m_0 
                                                              - 9 b_4 m_0 + 3 b_5 m_0 + 3 b_6 m_0 - 9 b_7 m_0\nonumber\\
       & &\qquad \qquad \quad                                 - 12 b_9 m_0 + 9 b_{15} m_0^2 - 3 b_{16} m_0^2 
                                                     + 9 b_{17} m_0^2 + 3 b_{18} m_0^2 + 12 b_{19} m_0^2\bigg). \nonumber\\
   \epsilon_{2,N,\eta}^{\pi\pi} & = &\frac{1}{54(4\pi F_0)^2}\bigg(-36 b_1 + 12 b_2- 4 b_3 - 24 b_8 + 36 b_D - 60 b_F 
                                                                   + 24 b_0 - \frac{D^2}{m_0} + \frac{6 D F}{m_0}\nonumber\\
       & &\qquad \qquad \quad                                      -  \frac{9 F^2}{m_0} -  9 b_4 m_0 + 3 b_5 m_0 
                                                                  + 3 b_6 m_0 - b_7 m_0 - 6 b_9 m_0 + 9 b_{15} m_0^2\nonumber\\
       & &\qquad \qquad \quad                               - 3 b_{16} m_0^2 +  b_{17} m_0^2 + 6 b_{19} m_0^2\bigg),\nonumber\\
   \epsilon_{2,N,\eta}^{\pi K} & = & \frac{1}{54(4\pi F_0)^2}\bigg(288 b_1 - 96 b_2+ 32 b_3 + 192 b_8 - 240 b_D + 336 b_F 
                                                                   - 192 b_0 + \frac{8 D^2}{m_0}\nonumber\\
    & &\qquad \qquad \quad                                         - \frac{48 D F}{m_0} + \frac{72 F^2}{m_0}+ 72 b_4 m_0 
                                                                   - 24 b_5 m_0 - 24 b_6 m_0 + 8 b_7 m_0 \nonumber\\
    & &\qquad \qquad \quad                                         + 48 b_9 m_0 - 72 b_{15} m_0^2 + 24 b_{16} m_0^2  
                                                                   -  8 b_{17} m_0^2 - 48 b_{19} m_0^2\bigg),  \nonumber\\
   \epsilon_{2,N,\eta}^{KK} & = &\frac{1}{54(4\pi F_0)^2} \bigg(-576 b_1 + 192 b_2 - 64 b_3 - 384 b_8 + 384 b_D - 384 b_F 
                                                                + 384 b_0 - \frac{16 D^2}{m_0}\nonumber\\ 
       & &\qquad \qquad \quad                                   + \frac{96 D F}{m_0} -  \frac{144 F^2}{m_0} - 144 b_4 m_0 
                                                                + 48 b_5 m_0    + 48 b_6 m_0 - 16 b_7 m_0  \nonumber\\
       & &\qquad \qquad \quad                                   - 96 b_9 m_0 + 144 b_{15} m_0^2  -  48 b_{16} m_0^2 
                                                              + 16 b_{17} m_0^2 + 96 b_{19} m_0^2 \bigg)~. 
\end{eqnarray}
The corresponding coefficients of the $\Lambda$-dependent term are
\begin{eqnarray} 
   \beta_{1,N,\pi}  & = &  \frac{m_0}{ (4 \pi F_0)^2}\bigg(-3 b_4 - 3 b_5 - 3 b_6 - 3 b_7 - 6 b_9 + m_0(3 b_{15} +3 b_{16} + 3 b_{17}+ 6 b_{19})\bigg),    \nonumber\\
   \beta_{1,N,K}  & = &   \frac{m_0}{8 (\pi F_0 )^2}\bigg(-b_{12}-3 b_4+ b_5 + b_6 -3 b_7 -4 b_9 + m_0(3 b_{15} -  b_{16} + 3 b_{17} + b_{18} + 4 b_{19})\bigg),   \nonumber\\
   \beta_{1,N,\eta}  & = &  \frac{m_0}{3(4\pi F_0 )^2 }\bigg(-9 b_4 + 3 b_5 + 3 b_6 - b_7 - 6 b_9 + m_0 (9 b_{15} - 3 b_{16} +b_{17} + 6 b_{19})\bigg),    \nonumber\\
   \beta_{2,N,\pi}^\pi  & = & \frac{1}{(4\pi F_0 )^2 } \bigg(-6 b_1 - 6 b_2 - 6 b_3 - 12 b_8 + 6 b_D + 6 b_F 
                                                                + 12 b_0 -  3 b_4 m_0 - 3 b_5 m_0\nonumber\\
         & &\qquad \qquad \quad                                 - 3 b_6 m_0 - 3 b_7 m_0                                - 6 b_9 m_0 +  3 b_{15} m_0^2 + 3 b_{16} m_0^2 
                                                                +  3 b_{17} m_0^2 + 6 b_{19}m_0^2\bigg),    \nonumber\\ 
%
%
   \beta_{2,N,K}^\pi  & = &  \frac{1}{9 (4\pi F_0 )^2 } \bigg(-52 b_D D^2 + 60 b_F D^2 + 120 b_D D F - 72 b_F D F 
                                                              - 36 b_D F^2 + 108 b_F F^2\bigg),   \nonumber\\
   \beta_{2,N,K}^K  & = &   \frac{1}{9 (4\pi F_0 )^2 }\bigg(-108 b_1 + 36 b_2 - 108 b_3 - 144 b_8 + 108 b_D - 36 b_F 
                                                            + 144 b_0 + 52 b_D D^2\nonumber\\
           & &\qquad \qquad \quad                            -  60 b_F D^2 - 120 b_D D F + 72 b_F D F + 36 b_D F^2 
                                                             - 108 b_F F^2                      - 18 b_{12} m_0 \nonumber\\
            & &\qquad \qquad \quad                            - 54 b_4 m_0 + 18 b_5 m_0 + 18 b_6 m_0
                                                             - 54 b_7 m_0 - 72 b_9 m_0 + 54 b_{15} m_0^2\nonumber\\
           & &\qquad \qquad \quad                               -  18 b_{16} m_0^2 
                                                              + 54 b_{17} m_0^2 + 18 b_{18} m_0^2 + 72 b_{19} m_0^2\bigg),   \nonumber\\
   \beta_{2,N,\eta}^\pi  & = & \frac{1}{9 (4\pi F_0 )^2 }\bigg(18 b_1 - 6 b_2 + 2 b_3 + 12 b_8 - 18 b_D + 30 b_F 
                                                                - 12 b_0  + 9 b_4 m_0 - 3 b_5 m_0\nonumber\\ 
           & &\qquad \qquad \quad                               - 3 b_6 m_0 + b_7 m_0 + 6 b_9 m_0 
                                                                 - 9 b_{15} m_0^2 + 3 b_{16} m_0^2-  b_{17} m_0^2 
                                                                - 6 b_{19} m_0^2\bigg),\nonumber\\
     \beta_{2,N,\eta}^K  & = &  \frac{1}{9 (4\pi F_0 )^2 }\bigg(-72 b_1 + 24 b_2 - 8 b_3 - 48 b_8 + 48 b_D - 48 b_F 
                                                             + 48 b_0 -  36 b_4 m_0 \nonumber\\
            & &\qquad \qquad \quad                           + 12 b_5 m_0 + 12 b_6 m_0 - 4 b_7 m_0 - 24 b_9 m_0 
                                                             + 36 b_{15} m_0^2 - 12 b_{16} m_0^2\nonumber\\ 
           & &\qquad \qquad \quad                            +  4 b_{17} m_0^2 + 24 b_{19} m_0^2\bigg), \nonumber\\
     \beta_{3,N,\pi}^{\pi\pi}  & = & \frac{1}{2 (4\pi F_0 )^2 }\bigg (-12 b_1 - 12 b_2 - 12 b_3 - 24 b_8 + 12 b_D 
                                                                      + 12 b_F + 24 b_0 - \frac{3 D^2}{m_0}\nonumber\\  
            & &\qquad \qquad \quad                                    -\frac{6DF}{m_0}-\frac{3F^2}{m_o} -3 b_4 m_0 - 3 b_5 m_0 
                                                                      - 3 b_6 m_0 - 3 b_7 m_0\nonumber\\ 
           & &\qquad \qquad \quad                                     -6  b_9 m_0 + 3 b_{15} m_0^2
                                     + 3 b_{16} m_0^2 + 3 b_{17} m_0^2 + 6 b_{19} m_0^2\bigg),\nonumber\\
%
    \beta_{3,N,K}^{\pi K}  & = &\frac{1}{3 (4\pi F_0 )^2 }\bigg(-52 b_D D^2 + 60 b_F\ D^2 + 120 b_D\ D F - 72 b_F D F 
                                                                - 36 b_D F^2 + 108 b_F F^2\bigg),      \nonumber\\
    \beta_{3,N,K}^{K K}  & = &  \frac{1}{3 (4\pi F_0 )^2 } \bigg(-36 b_1 + 12 b_2 - 36 b_3 - 48 b_8 + 36 b_D - 12 b_F 
                                                                 + 48 b_0 + 52 b_D D^2\nonumber\\
           & &\qquad \qquad \quad                                -  60 b_F D^2 -  120 b_D D F + 72 b_F D F+ 36 b_D F^2 
                                                                 - 108 b_F F^2 - \frac{5 D^2}{m_0}\nonumber\\
           & &\qquad \qquad \quad                                + \frac{6 D F}{m_0} -  \frac{9 F^2}{m_0} - 3 b_{12} m_0 
                                                                 - 9 b_4 m_0 +  3 b_5 m_0 + 3 b_6 m_0\nonumber\\
           & &\qquad \qquad \quad                                - 9 b_7 m_0 - 12 b_9 m_0 + 9 b_{15} m_0^2 - 3 b_{16} m_0^2 
                                                                  + 9 b_{17} m_0^2\nonumber\\
           & &\qquad \qquad \quad                                 + 3 b_{18}{m_0}^2 + 12 b_{19} m_0^2\bigg),       \nonumber\\
    \beta_{3,N,\eta}^{\pi\pi}  & = &  \frac{1}{ (4\pi F_0 )^2 }\bigg(-\frac{2 b_1}{3} + \frac{2 b_2}{9} - \frac{2 b_3}{27} 
                                                                     - \frac{4 b_8}{9} + \frac{2 b_D}{3} - \frac{10 b_F}{9} 
                                                                     + \frac{4 b_0}{9} - \frac{D^2}{54 m_0}\nonumber\\
          & &\qquad \qquad \quad                                     +  \frac{D F}{9 m_0} - \frac{F^2}{6 m_0} 
                                                                     - \frac{b4 m_0}{6} + \frac{b_5 m_0}{18} 
                                                                     + \frac{b_6 m_0}{18} - \frac{b_7 m_0}{54}\nonumber\\
          & &\qquad \qquad \quad                                      - \frac{b_9 m_0}{9} + \frac{b_{15} m_0^2}{6} 
                                                                     -  \frac{b_{16} m_0^2}{18} + \frac{b_{17} m_0^2}{54} 
                                                                     + \frac{b_{19} m_0^2}{9}\bigg),  \nonumber\\
    \beta_{3,N,\eta}^{\pi K}  & = &  \frac{1}{ (4\pi F_0 )^2 }\bigg(\frac{16 b_1}{3} - \frac{16 b_2}{9} + \frac{16 b_3}{27} 
                                                                    + \frac{32 b_8}{9} - \frac{40 b_D}{9} +\frac{56 b_F}{9} 
                                                                    - \frac{32 b_0}{9} + \frac{4 D^2}{27 m_0}\nonumber\\
          & &\qquad \qquad \quad                                    - \frac{8 D F}{9 m_0} + \frac{4 F^2}{3 m_0} 
                                                                    +  \frac{4 b_4 m_0}{3} - \frac{4 b_5 m_0}{9} 
                                                                    - \frac{4 b_6 m_0}{9} + \frac{4 b_7 m_0}{27}\nonumber\\
         & &\qquad \qquad \quad                                     + \frac{8 b_9 m_0}{9} - \frac{4 b_{15} m_0^2}{3} 
                                                                    + \frac{4 b_{16} m_0^2}{9} - \frac{4 b_{17} m_0^2}{27} 
                                                                    - \frac{8 b_{19} m_0^2}{9}\bigg),   \nonumber\\
    \beta_{3,N,\eta}^{K K}  & = &   \frac{1}{ (4\pi F_0 )^2 }\bigg(-\frac{32 b_1}{3}    + \frac{32 b_2}{9} - \frac{32 b_3}{27}
                                                                   - \frac{64 b_8}{9} + \frac{64 b_D}{9} - \frac{64 b_F}{9} 
                                                                   + \frac{64 b_0}{9} - \frac{8 D^2}{27 m_0} \nonumber\\
       & &\qquad \qquad \quad                                      + \frac{16 D F}{9 m_0} - \frac{8 F^2}{3 m_0} 
                                                                   -  \frac{8 b_4 m_0}{3} + \frac{8 b_5 m_0}{9} 
                                                                   + \frac{8 b_6 m_0}{9} - \frac{8 b_7 m_0}{27}\nonumber\\ 
        & &\qquad \qquad \quad                                     - \frac{16 b_9 m_0}{9} + \frac{8 b_{15} m_0^2}{3} 
                                                                   - \frac{8 b_{16} m_0^2}{9} + \frac{8 b_{17} m_0^2}{27} 
                                                                   + \frac{16 b_{19} m_0^2}{9}\bigg)~.   
\end{eqnarray}
Finally, we note that we use the values of the $b_i$ as given in
Table~\ref{LECb} and set all other dimension two LECs to zero.

\def\theequation{\Alph{section}.\arabic{equation}}
\setcounter{equation}{0}
\section{Loop integrals in cut-off regularization}
\label{app:int}

In this appendix, we collect all integrals needed in the calculation of the
baryon self-energy to fourth order utilizing cut-off regularization. 
Here, $M$ is a generic symbol for the propagating Goldstone boson and we 
give the explicit representations only for the relevant case $M > |\omega|$:
\begin{eqnarray}
    I_1(M,\omega) & \doteq & \int \frac{\mbox{d}^4k}{(2\pi)^4} 
      \frac{i}{k^2-M^2+i0^+}\frac{i}{\omega-v \cdot k+i0^+}
      (S\cdot k)^2   \nonumber\\
       &  = & -i  \frac{1}{16\pi^2} 
      \Bigg\{\frac{\Lambda^3}{3}-(M^2-\omega^2)\Lambda
      +\bigg(M^2-\omega^2\bigg)^\frac{3}{2}\pi
      -\bigg(M^2-\omega^2\bigg)^\frac{3}{2} 
       \arctan\bigg(\frac{\sqrt{M^2-\omega^2}}{\Lambda}\bigg)\nonumber\\
       &  & \quad\quad\quad\quad+\frac{\omega\Lambda^2}{2}\sqrt{1+\frac{M^2}{\Lambda^2}}
       +\bigg(\omega^3-\frac{3}{2}\omega M^2\bigg) 
       \ln\bigg(\frac{\Lambda}{M}+\sqrt{1+\frac{\Lambda^2}{M^2}}\bigg)
\nonumber\\&&    \quad\quad\quad\quad    -(M^2-\omega^2)^\frac{3}{2}\arctan\bigg( \sqrt{\frac{M^2}{\omega^2}-1}
       \sqrt{1+\frac{M^2}{\Lambda^2}}\bigg)
       \Bigg\}~,  \nonumber\\
  I_1(M,0) & = & -i  \frac{1}{16\pi^2} 
      \Bigg\{\frac{\Lambda^3}{3}-M^2\Lambda
    + M^3\frac{\pi}{2}-M^3 \arctan\bigg(\frac{M}{\Lambda}\bigg)\Bigg\}~,\\\nonumber\\\nonumber\\
    I_2(M,\omega) & \doteq &  \int \frac{\mbox{d}^4k}{(2\pi)^4}
       \frac{i}{k^2-M^2+i0^+}\frac{i}{\omega-v \cdot k+i0^+}
       (S\cdot k)^2 (v\cdot k)   \nonumber\\
       & = & -i\frac{1}{16 \pi^2}\Bigg\{\frac{1}{4}\Lambda^4
       \sqrt{1+\frac{M^2}{\Lambda^2}}-\frac{3}{8} M^2\Lambda^2 
       \sqrt{1+\frac{M^2}{\Lambda^2}}
       +\frac{3}{8} M^4\ln\bigg(\frac{\Lambda}{M}+\sqrt{1+\frac{\Lambda^2}{M^2}}\bigg)\Bigg\}
       +\omega I_1(M,\omega)~,\nonumber\\
    I_2(M,0) & = & -i\frac{1}{16 \pi^2}\Bigg\{\frac{1}{4}\Lambda^4
       \sqrt{1+\frac{M^2}{\Lambda^2}}-\frac{3}{8} M^2\Lambda^2 
       \sqrt{1+\frac{M^2}{\Lambda^2}}
\nonumber\\ 
       &   & \quad\quad\quad\quad
+\frac{3}{8} M^4 \bigg[\ln\bigg(\frac{\Lambda}{m_0}\bigg)-\ln\bigg(\frac{M}{m_0}\bigg)
       +\ln\bigg(1+\sqrt{1+\frac{M^2}{\Lambda^2}}\bigg)\bigg]\bigg\}~,\nonumber\\
    I_3(M,\omega) & \doteq &  \int \frac{\mbox{d}^4k}{(2\pi)^4}
       \frac{i}{k^2-M^2+i0^+}\frac{i^2}{(\omega-v \cdot k+i0^+)^2}
       (S\cdot k)^2  
        =  -i \frac{\mbox{d}}{\mbox{d}\omega}I_1(M,\omega)\nonumber\\
        & = & -\frac{1}{16 \pi^2}\Bigg\{\frac{3}{2}\Lambda^2
        \sqrt{1+\frac{M^2}{\Lambda^2}}+3\omega \Lambda-\frac{\Lambda^2}{\Lambda^2+M^2-\omega^2}\bigg(\omega \Lambda+ \Lambda^2 \sqrt{1+\frac{M^2}{\Lambda^2}}\bigg)\nonumber\\&& \quad\quad\quad\quad-3\omega \sqrt{M^2- \omega^2}\pi +\bigg(3\omega^2-\frac{3}{2}M^2\bigg)  \ln\bigg(\frac{\Lambda}{M}+\sqrt{1+\frac{\Lambda^2}{M^2}}\bigg)\nonumber\\&& \quad\quad\quad\quad
         +3  \omega \sqrt{M^2- \omega^2}\arctan\bigg(\frac{\sqrt{M^2-\omega^2}}{\Lambda}\bigg)\nonumber\\&& \quad\quad\quad\quad+3  \omega \sqrt{M^2- \omega^2} \arctan\bigg(\sqrt{\frac{M^2}{\omega^2}-1}\sqrt{1
         +\frac{M^2}{\Lambda^2}}\bigg)\Bigg\}~,\nonumber\\
    I_3(M,0) & = & -\frac{1}{16\pi^2}\Bigg\{\frac{\Lambda^2}{2}
       \sqrt{1+\frac{M^2}{\Lambda^2}}+M^2 
        \frac{1}{\sqrt{1+\frac{M^2}{\Lambda^2}}}
      \nonumber\\&& \quad\quad\quad\quad  -\frac{3}{2} M^2 \bigg[\ln\bigg(\frac{\Lambda}{m_0}\bigg)-\ln\bigg(\frac{M}{m_0}\bigg)+
        \ln\bigg(1+\sqrt{1+\frac{M^2}{\Lambda^2}}\bigg)\bigg]\Bigg\}~, \nonumber\\
I_4(M,\omega) & \doteq &  \int \frac{\mbox{d}^4k}{(2\pi)^4}
       \frac{i}{k^2-M^2+i0^+}\frac{i^2}{(\omega-v \cdot k+i0^+)^2}
       (S\cdot k)^2  (v\cdot k) 
        =  -i \frac{\mbox{d}}{\mbox{d}\omega}I_2(M,\omega)\nonumber\\
        & = &-i I_1(M,\omega)+\omega I_3(M,\omega)~, \nonumber\\
    I_4(M,0) & = & -i I_1(M,0)~, \nonumber\\
%
    I_5(M,\omega) & \doteq &  \int \frac{\mbox{d}^4k}{(2\pi)^4}
       \frac{i}{k^2-M^2+i0^+}\frac{i^2}{(\omega-v \cdot k+i0^+)^2}
       (S\cdot k)^2 (v\cdot k)^2   \nonumber\\
      & = & -i\omega I_1(M,\omega) -i  I_2(M,\omega)+\omega^2 I_3(M,\omega)~, \nonumber\\
    I_5(M,0) & = & -i I_2(M,0)~, \nonumber\\
    I_6(M,\omega) & \doteq &  \int \frac{\mbox{d}^4k}{(2\pi)^4}
       \frac{i}{k^2-M^2+i0^+}\frac{i^2}{(\omega-v \cdot k+i0^+)^2}
       (S\cdot k)^2  k^2  
       =  M^2 I_3(M,\omega)~,\nonumber\\
    I_6(M,0) & = &  M^2 I_3(M,0)~.
\end{eqnarray}

\noindent Similarly, for the calculation of the baryon tadpole and the meson masses in CR we need the following
integrals:
\begin{eqnarray}
    \alpha_1(M) & \doteq & \int \frac{\mbox{d}^4k}{(2\pi)^4} \frac{i}{k^2-M^2+i0^+}
        \nonumber\\
        &  = & \frac{1}{2(2\pi)^2} \left\{ \Lambda^2\sqrt{1+\frac{M^2}
        {\Lambda^2}} - M^2 \left[\ln\bigg(\frac{\Lambda}{m_0}\bigg)
        -\ln\bigg(\frac{M}{m_0}\bigg)
        +\ln\bigg(1+\sqrt{1+\frac{M^2}{\Lambda^2}}\bigg)\right]\right\}~,\nonumber\\
%
    \alpha_2(M) & \doteq & \int \frac{\mbox{d}^4k}{(2\pi)^4} \frac{i}{k^2-M^2+i0^+}
         k^2
          =  M^2 \alpha_1(M)~,\nonumber\\
    \alpha_3(M) & \doteq & \int \frac{\mbox{d}^4k}{(2\pi)^4} \frac{i}{k^2-M^2+i0^+}
        k_0^2
         =  \frac{1}{(2\pi)^2} \frac{1}{4}\Lambda^4 
         \sqrt{1+\frac{M^2}{\Lambda^2}}+\frac{1}{4}
         M^2\alpha_1(M)~.
      \nonumber 
\end{eqnarray}

\def\theequation{\Alph{section}.\arabic{equation}}
\setcounter{equation}{0}
\section{Meson masses in cut-off regularization}
\label{app:mesonmasses}

Here, we collect the formualae for the meson masses to fourth order in CR. The
pertinent diagrams are tree graphs with one insertion from  
${\cal L}_\phi^{(2)}$ and ${\cal L}_\phi^{(4)}$ and tadpoles 
with exactly one insertion from ${\cal L}_\phi^{(2)}$.
We have
\begin{eqnarray}
    M_\pi^2  & = & M_{0,\pi}^2+\delta M_\pi^{(4)}\nonumber\\
             & = & M_{0,\pi}^2+\frac{M_\pi^4}{16 F_0^2 \pi^2}\ln\frac{M_\pi}{m_0}
                 -\frac{M_\pi^2 M_\eta^2}{48 F_0^2 \pi^2}\ln\frac{M_\eta}{m_0}
                 -\frac{16 L_4^{(r)} M_K^2 M_\pi^2}{F_0^2} 
               + \frac{32 L_6^{(r)} M_K^2 M_\pi^2}{F_0^2}
               - \frac{8 L_4^{(r)} M_\pi^4}{F_0^2}
                 -\frac{8 L_5^{(r)} M_\pi^4}{F_0^2} \nonumber\\
&+& \frac{16 L_6^{(r)} M_\pi^4}{F_0^2} + \frac{16 L_8^{(r)} M_\pi^4}{F_0^2}
          -\frac{M_\pi^4}{16 F_0^2 \pi^2}\ln\Bigg(1+\sqrt{1
           +\bigg(\frac{M_\pi}{\Lambda}\bigg)^2}\Bigg)
                 +\frac{M_\pi^2 M_\eta^2}{48 F_0^2 \pi^2}\ln\Bigg(1
                 +\sqrt{1+\bigg(\frac{M_\eta}{\Lambda}\bigg)^2}\Bigg)\nonumber\\
          &+&     \frac{1}{\sqrt{1+\big(\frac{M_\pi}{\Lambda}}\big)^2}
                 \Bigg\{\frac{M_\pi^2}{16 F_0^2 \pi^2} 
                  \Lambda^2\bigg(1-\sqrt{1+\bigg(\frac{M_\pi}{\Lambda}}\bigg)^2\bigg)+\frac{M_\pi^4}{16 F_0^2 \pi^2}\Bigg\}\nonumber\\        
          &+&\frac{1}{\sqrt{1+\big(\frac{M_\eta}{\Lambda}\big)^2}}
                      \Bigg\{-\frac{M_\pi^2}{48 F_0^2 \pi^2} 
                  \Lambda^2\bigg(1-\sqrt{1+\bigg(\frac{M_\eta}{\Lambda}\bigg)^2}\bigg)-\frac{M_\pi^2 M_\eta^2}
                  {48 F_0^2 \pi^2}\Bigg\},
\end{eqnarray}
 
\begin{eqnarray}
    M_K^2   & = & M_{0,K}^2+\delta M_K^{(4)}\nonumber\\ 
            & = & M_{0,K}^2+\frac{M_K^2 M_\eta^2}{24 F_0^2 \pi^2}\ln\frac{M_\eta}{m_0}
            +\frac{ M_K^2 M_\pi^2}{F_0^2}(-8 L_4^{(r)}+16 L_6^{(r)}) 
            +\frac{ M_K^4}{F_0^2}(-16 L_4^{(r)}-8 L_5^{(r)}+32 L_6^{(r)}+16        
            L_8^{(r)})\nonumber\\   
            &-&\frac{M_K^2 M_\eta^2}{24 F_0^2 \pi^2}\ln\Bigg(1
            +\sqrt{1+\bigg(\frac{M_\eta}{\Lambda}\bigg)^2}\Bigg)
            +\frac{1}{\sqrt{1+\big(\frac{M_\eta}{\Lambda}\big)^2}}
                      \Bigg\{\frac{M_K^2}{24 F_0^2 \pi^2} 
           \Lambda^2\bigg(1-\sqrt{1+\bigg(\frac{M_\eta}{\Lambda}\bigg)^2}\bigg)
           -\frac{M_K^2 M_\eta^2}{24 F_0^2 \pi^2}\Bigg\}~,\nonumber\\&& 
\end{eqnarray}
\begin{eqnarray}
    M_\eta^2 & = & M_{0,\eta}^2+\delta M_\eta^{(4)}\nonumber\\
             & = & M_{0,\eta}^2-\frac{M_\pi^4}{16 F_0^2 \pi^2}\ln\frac{M_\pi}{m_0}
                 +\frac{M_K^4}{6 F_0^2 \pi^2}\ln\frac{M_K}{m_0}
                 +\Big(-\frac{7M_\pi^4}{432 F_0^2 \pi^2}
                 +\frac{11 M_K^2 M_\pi^2}{108 F_0^2\pi^2}
                 -\frac{4 M_K^4}{27 F_0^2\pi^2}\Big)\ln\frac{M_\eta}{m_0}\nonumber\\
             &&    +\frac{M_K^4} {F_0^2}(-\frac{64 L_4^{(r)}}{3} - \frac{128 L_5^{(r)}}{9} + \frac{128 L_6^{(r)}}{3} 
                 + \frac{128 L_7^{(r)}}{3} + \frac{128 L_8^{(r)}}{3} )\nonumber\\
             &&    + \frac{M_K^2 M_\pi^2}{F_0^2} (-\frac{16 L_4^{(r)}}{3} + \frac{64 L_5^{(r)}}{9} + \frac{32 L_6^{(r)}}{3}
                 - \frac{256 L_7^{(r)}}{3} - \frac{128 L_8^{(r)}}{3})\nonumber\\
             &&    + \frac{M_\pi^4}{F_0^2}(\frac{8 L_4^{(r)}}{3} - \frac{8 L_5^{(r)}}{9} - \frac{16 L_6^{(r)}}{3} 
                 + \frac{128 L_7^{(r)}}{3} + 16 L_8^{(r)}) \nonumber\\
             &&     +\frac{M_\pi^4}{16 F_0^2 \pi^2}\ln\Bigg(1+\sqrt{1+\bigg(\frac{M_\pi}{\Lambda}\bigg)^2}\Bigg)
              -\frac{M_K^4}{6 F_0^2 \pi^2}\ln\Bigg(1+\sqrt{1
               +\bigg(\frac{M_K}{\Lambda}\bigg)^2}\Bigg)\nonumber\\
             &&     -\big(-\frac{7M_\pi^4}{432 F_0^2 \pi^2}+\frac{11 M_K^2 M_\pi^2}{108 F_0^2\pi^2}
                 -\frac{4 M_K^4}{27 F_0^2\pi^2}\big)\ln\Bigg(1+\sqrt{1+\bigg(\frac{M_\eta}{\Lambda}\bigg)^2}\Bigg)\nonumber\\
          &&     +\frac{1}{\sqrt{1+\big(\frac{M_\pi}{\Lambda}}\big)^2}\Bigg\{-\frac{M_\pi^2}{16 F_0^2 \pi^2} 
                  \Lambda^2\bigg(1-\sqrt{1+\bigg(\frac{M_\pi}{\Lambda}}\bigg)^2\bigg)
                 -\frac{M_\pi^4}{16 F_0^2 \pi^2}\Bigg\}\nonumber\\
         &&      +\frac{1}{\sqrt{1+\big(\frac{M_K}{\Lambda}\big)^2}}
                      \Bigg\{\frac{M_K^2}{6 F_0^2 \pi^2} 
                  \Lambda^2\bigg(1-\sqrt{1+\bigg(\frac{M_K}{\Lambda}\bigg)^2}\bigg)+\frac{M_K^4}
                  {6 F_0^2 \pi^2}\Bigg\}\\
         &&         +\frac{1}{\sqrt{1+\big(\frac{M_\eta}{\Lambda}\big)^2}}
                      \Bigg\{\big(-\frac{M_K^2}{9 F_0^2 \pi^2}+\frac{7 M_\pi^2}{144 F_0^2\pi^2}) 
               \Lambda^2\bigg(1-\sqrt{1+\bigg(\frac{M_\eta}{\Lambda}\bigg)^2}\bigg)
               -\frac{4 M_K^2}{27 F_0^2\pi^2}
               +\frac{11 M_\pi^2 M_K^2}{108 F_0^2 \pi^2}
               -\frac{7 M_\pi^2}{432 F_0^2\pi^2}\Bigg\}~.\nonumber
\end{eqnarray}
As required, in the limit $\Lambda \to \infty$ we recover the standard DR
result \cite{GL85}. The polynomial and logarithmic divergences in the cut-off
are taken care of by the following renormalization (note again that e.g. $B_0$
is not renormalized in DR):
\begin{eqnarray}
   B_0^{(r)} &= & B_0+\frac{1}{24 \pi^2 F_0^2}B_0\Lambda^2 ~, \quad
   L_7^{(r)}+\frac{L_8^{(r)}}{3} = L_7+\frac{L_8}{3} +\frac{5}{2304\pi^2}\ln\frac{\Lambda}{m_0},\nonumber\\ 
   L_5^{(r)}- 2 L_8^{(r)}&=& L_5 - 2 L_8 +\frac{1}{96 \pi^2}\ln\frac{\Lambda}{m_0}~,
   L_4^{(r)} - 2 L_6^{(r)} = L_4 - 2 L_6 -\frac{ 1}{576 \pi^2}\ln\frac{\Lambda}{m_0}.
\end{eqnarray}
Here, $B_0$ connects the leading terms in the chiral expansion of the
Goldstone boson masses with the quark masses,
\begin{eqnarray}
   M_{0,\pi}^2=2 B_0^{(r)} \hat{m}~,\quad
   M_{0,K}^2 =  B_0^{(r)} (\hat{m} + m_s)~,\quad
   M_{0,\eta}^2 = \frac{2}{3}B_0^{(r)}(\hat{m} + 2 m_s)~,
\end{eqnarray}
with $\hat m$ the average light quark mass.

\pagebreak

\pagebreak

\section*{Figures}

\vspace{1.5cm}
\noindent

\pagebreak

$\,$

\vspace{1cm}

\begin{figure}[htb]
\vspace{0.9cm}
\parbox{.50\textwidth}{\epsfig{file= mNp4.eps,width=.46\textwidth,angle=0,
silent=,clip=}}
\hfill
\parbox{.50\textwidth}{\epsfig{file= mNk4.eps,width=.46\textwidth,angle=0,
silent=,clip=}}
\vspace{0.2cm}
\begin{center}
\caption{Left panel: Pion mass dependence of the nucleon mass for various
sets of the LECs $d_i$ as explained in the text. Right panel: Kaon mass
dependence.
\label{fig:mN4}}
\end{center}
\end{figure}
\noindent

\pagebreak

\noindent

\end{document}